\documentclass{aastex63} 
\usepackage{verbatim}
\usepackage{float}
\usepackage{graphicx}
\usepackage{epstopdf}
\usepackage{subfigure}
\usepackage{CJK}
\usepackage{amsmath}

\usepackage{booktabs}
\usepackage{url}
\usepackage{amssymb}
\makeatletter

\newcommand{\Rmnum}[1]{\expandafter\@slowromancap\romannumeral #1@}
\makeatother

\shorttitle{}
\shortauthors{}

\begin{document}
\begin{CJK*}{UTF8}{gbsn}

\title{Gaussian Process Modeling Blazar Multiwavelength Variability: 
Indirectly Resolving Jet Structure}

\correspondingauthor{Dahai Yan}
\email{yandahai@ynu.edu.cn}
\correspondingauthor{Li Zhang}
\email{lizhang@ynu.edu.cn}

\author{Haiyun Zhang (张海云)}
\affiliation{Department of Astronomy, Key Laboratory of Astroparticle Physics of Yunnan Province, Yunnan University, Kunming 650091, China}

\author{Dahai Yan (闫大海)}
\affiliation{Department of Astronomy, Key Laboratory of Astroparticle Physics of Yunnan Province, Yunnan University, Kunming 650091, China}

\author{Li Zhang (张力)}
\affiliation{Department of Astronomy, Key Laboratory of Astroparticle Physics of Yunnan Province, Yunnan University, Kunming 650091, China}

\begin{abstract}
Blazar jet structure can be indirectly resolved by analyzing the multiwavelength variability.
In this work, we analyze the long-term variability of blazars in radio, optical and X-ray energies with the Gaussian process (GP) method.
The multiwavelength variability can be successfully characterized by the damped-random walk (DRW) model.
The nonthermal optical characteristic timescales 
of 38 blazars are statistically consistent with the $\gamma$-ray characteristic timescales of 22 blazars. 
For three individuals (3C 273, PKS 1510-089, and BL Lac), the nonthermal optical, X-ray, and $\gamma$-ray characteristic timescales are also consistent within the measured 95$\%$ errors, but the radio timescale of 3C 273 is too large to be constrained by the decade-long light curve.
The synchrotron and inverse-Compton emissions have the same power spectral density, 
suggesting that the long-term jet variability is irrelevant to the emission mechanism.
In the plot of the rest-frame timescale versus black hole mass, 
the optical-$\gamma$-ray timescales of the jet variability occupy almost the same space with the timescales of accretion disk emission from normal quasars, 
which may imply that the long-term variabilities of the jet and accretion disk are driven by the same physical process.
It is suggested that the nonthermal optical-X-ray and $\gamma$-ray emissions are produced in the same region,
while the radio core which can be resolved by very-long-baseline interferometry locates at a far more distant region from the black hole.
Our study suggests a new methodology for comparing thermal and nonthermal emissions, which is achieved by using the standard GP method.

\end{abstract}   

\section{Introduction} \label{sec:intro}

Flat spectrum radio quasars (FSRQs) and BL Lac objects (BL Lacs)  are classed into a special class of active galactic nuclei (AGNs) called blazars, whose jets nearly point to the Earth.
Blazars are highly variable over the entire electromagnetic bands.  
One popular scenario is that the accretion onto a supermassive black hole is the central engine, 
driving relativistic jet. 
But the detailed process is still unclear.
Thanks to the high variability of blazars, one can investigate the physical process close to the central engine \citep[e.g.,][]{2019Galax...7...28R}, 
such as the location of the emitting region and the jet-disk connection \citep[e.g.,][]{2016ApJ...824L..20A,2019ApJ...877...39M,2022ApJ...930..157Z}.

Using advanced interferometric instruments, blazar radio jet can be directly resovled on $\sim$parsec-scale \citep[see][for a recent review]{2019NewAR..8701541H}. 
This provides a calibrator for multi-band variability analysis.
There have been lots of works attempting to investigate the underlying physical process of blazar jet with multi-band variability \citep[e.g.,][]{2012ApJ...749..191C,2013ApJ...773..177N,2017ApJS..229...21X,2018ApJ...863..175G,2022ApJ...927..214G}. 
\cite{2014MNRAS.445..428M} investigated the time-domain relationship between radio and $\gamma$-ray emission of blazars, 
and found the correlations only exist in a minority of the sources over a 4 yr period. 
They found radio variations lagging the $\gamma$-ray variations, 
suggesting that the $\gamma$-ray emission originates upstream of the radio emission.
This result is further verified by \cite{2018MNRAS.480.5517L} who concluded that the radio variation is usually substantially delayed to the other wavelengths for blazars.
\citet{2021ApJ...923....7B} analyzed the correlation between optical ($V$-band) and $\gamma$-ray variabilities for blazars and found that the optical variability is highly 
correlated with the $\gamma$-ray variability except for 3C 273, however, no significant lagging is found.
The multi-band variability analysis can be considered as an indirect approach for resolving blazar jet. 

The GP method becomes popular in modern time-domain astronomy \citep[e.g.,][]{2019ApJ...885...12R,2021Sci...373..789B,2021ApJ...907..105Y,2021ApJ...914..144G,2022MNRAS.513.2841C, 2022ApJ...934....6R,2022MNRAS.514..164S,2022ApJ...930..157Z}.
The GP method enables us to effectively extract information from astronomical variability.
For example, \citet{2022ApJ...930..157Z} used a GP method to characterize the $\gamma$-ray variability of AGNs with stochastic process.
It is found that the DRW model can successfully fit the $\gamma$-ray variability, 
which is similar with the optical variability of AGN accretion disk \citep{2009ApJ...698..895K,2018MNRAS.476L..55L,2021Sci...373..789B}.
Moreover, \citet{2022ApJ...930..157Z} suggested a connection between the jet and the accretion disk by comparing the rest-frame
$\gamma$-ray timescales of blazars with the optical accretion disk timescales of quasars.
 
In this work, we analyze the multi-band variability of blazars with the GP method, which is independent of the temporal correlation analysis. 
We hope to extract additional information from the variability.
Using the data from Fermi-Large Area Telescope (Fermi-LAT), we carried out systematic research of $\gamma$-ray variability of AGNs recently \citep{2022ApJ...930..157Z}.
So far, the Small and Moderate Aperture Research Telescope System (SMARTS) monitoring program\footnote{\url{http://www.astro.yale.edu/smarts/glast/home.php}} \citep{2012ApJ...756...13B} and the Steward Observatory (SO) spectropolarimetric
monitoring project\footnote{\url{http://james.as.arizona.edu/~psmith/Fermi/datause.html}} \citep{2009arXiv0912.3621S} 
can provide almost ten years' (from 2008 to 2018) optical data of Fermi blazars.
RXTE AGN Timing $\&$ Spectral Database\footnote{\url{https://cass.ucsd.edu/~rxteagn/}} \citep{2013ApJ...772..114R} provides long-term X-ray data, 
and the Owens Valley Radio Observatory (OVRO) 40 m program \citep{2011ApJS..194...29R} provides radio light curves (LCs) from 2008 to 2020\footnote{\url{http://astro.caltech.edu/ovroblazars/}}.
Using these public data, we analyze the radio, optical and X-ray variability of three individual blazars, as well as optical variability for a sample including 38 Fermi blazars.
The format of this paper is as follows. 
In Section~\ref{sec:source and method}, we describe the data as well as the GP method. 
The modeling results of the three individual sources and 38 blazars are shown in Section~\ref{sec:results}.
We give discussions and physical interpretations of the results in Section~\ref{sec:discussion}.  
In Section~\ref{sec:summary}, we conclude the paper with a brief summary.

\section{Data and Gaussian Process Method} \label{sec:source and method}
\subsection{Data and Sources} \label{subsec:sample}
We use photometric data of blazars from the SMARTS and SO monitoring projects.
The SMARTS program gives photometric data at five wavelength bands
($B, V, R, J, K$), which were taken from the 1.3 m telescope at the Cerro Tololo Inter-American Observatory.
SO is a long-term optical program to support the Fermi Telescope, utilizing both the 2.3 m Bok Telescope on Kitt Peak and the 1.54 m Kuiper Telescope on Mt.Bigelow in Arizona.
The campaign of the SO program spanned almost a decade from 2008 November to 2018 July.
The X-ray data can be gained from RXTE observation which provided us with 16 yr (1996-2012) data in 2-10 keV.
OVRO 40 m program gives radio data of blazars from 2008 to 2020, which is a large-scale, fast-cadence 15 GHz radio monitoring program. 
We select sources having long-term continuous observations and a good sampling. 
For the source with a large gap in the LC, we only use the data covering a longer period before or after the gap for analysis.
Finally, We have 38 blazars in the optical band, including 23 FSRQs and 15 BL Lacs. 
Three blazars (3C 273, BL Lac, and PKS 1510-089) have long-term RXTE X-ray data. They are also in the sample of selected optical sources.
Unfortunately, among the three sources,  only 3C 273 has the OVRO LC. 
Table~\ref{tab:information} gives the general information of these targets.
 
\begin{deluxetable*}{ccccc}
	\tablecaption{Information of 38 blazars.\label{tab:information}}
	\tablewidth{0pt}
	\setlength{\tabcolsep}{4mm}{
	\tablehead{
		\colhead{Object} &  \colhead{$z$} & \colhead{Type} & \colhead{$logM_{\rm BH}/M_{\rm \odot}$} & \colhead{Ref.} 
	}
	\decimalcolnumbers
	\startdata
	1ES 1959+650 & 0.048 & BLL & $8.2\pm0.17$ & 1 \\
	1ES 2344+514 & 0.044 & BLL & $8.7\pm0.18$ & 2  \\
	3C 66A & 0.37 & BLL & $8.57^{0.03}_{0.6}$ & 3,4  \\
	3C 454.3 & 0.859 & FSRQ & $9.1\pm0.5$ & 6 \\
	PKS 0235+164 & 0.94 & BLL & $9.0$ & 7 \\
	4C +38.41 & 1.81396 & FSRQ & $9.5\pm0.5$ & 7 \\
	CTA 102 & 1.032 & FSRQ & 8.7 & 7 \\
	Mkn 421 & 0.03002 & BLL & $8.3\pm0.2$ & 6 \\
	Mkn 501 & 0.03298 & BLL & $9.2\pm0.2$ & 6 \\
	OJ 287 & 0.3056 & BLL & $8.8\pm0.5$ & 6 \\
	4C +21.35 & 0.43383 & FSRQ & $8.9\pm0.15$ & 8 \\
	PKS 2155-304 & 0.1167 & BLL & 8.9 & 9 \\
	S5 0716+714 & 0.31 & BLL & 8.7 & 10 \\
	W Com & 1.25813 & BLL & 8.5 & 14 \\
	4C +01.02 & 2.099 & FSRQ & 9.5 & 16 \\
	PKS 0208-512 & 1.003 & FSRQ & 9.2 & 7 \\
	PKS 0235-618 & 0.46657 & FSRQ & 9.0 & 14 \\ 
	PKS 0402-362 & 1.42284 & FSRQ & 9.0 & 14 \\
	PKS 0426-380 & 1.105 & BLL & 8.6 & 7 \\
	PKS 0458-02 & 2.286 & FSRQ & 8.7 & 11 \\
	PKS 0502+049 & 0.954 & FSRQ & $8.9\pm0.5$ & 12 \\
	PKS 0528+134 & 2.06 & FSRQ & 9.0 & 13 \\
	PMN J0531-4827 & 0.8116 & BLL & $\cdots$ & $\cdots$ \\
	PMN J0850-1213 & 0.566 & FSRQ & 8.7 & 14 \\
	PKS 1144-379 & 1.048 & BLL & 8.5 & 7 \\
	PKS 1244-255 & 0.633 & FSRQ & 8.3 & 14 \\
	PKS B1406-076 & 1.494 & FSRQ & 9.4 & 17 \\
	PKS 1730-130 & 0.902 & FSRQ & 8.7 & 14 \\
	PKS 1954-388 & 0.63 & FSRQ & $8.0\pm0.5$ & 5 \\
	PKS 2142-75 & 1.139 & FSRQ & 9.7 & 15 \\
	PKS 2233-148 & 0.33 & BLL & 8.7 & 14 \\
	PKS 2326-502 & 0.518 & FSRQ & 9.3 & 14 \\
	PMN J2345-1555 & 0.621 & FSRQ & $8.2\pm0.17$ & 8 \\
    Ton 599 & 0.72474 & FSRQ & $8.5\pm0.5$ & 5 \\
    PKS 2052-47 & 1.489 & FSRQ & 8.9 & 14 \\
    \hline
    3C 273 & 0.15834 & FSRQ & $8.9\pm0.5$ & 5 \\
    BL Lac & 0.0686 & BLL & $8.5\pm0.2$ & 6 \\
    PKS 1510-089 & 0.36 & FSRQ & $7.8^{+0.05}_{-0.04}$ & 18 \\
	\enddata 
	\tablecomments{(1) source name, (2) redshift, (3) source type, (4) black hole mass (in solar mass) collected from the references in the last column. 
	X-ray variability analysis is performed for the last three sources.
	  References: (1) \cite{2003ApJ...595..624F}, (2) \cite{2005ApJ...631..762W}, (3) \cite{2017MNRAS.469.2305K}, (4) \cite{2012NewA...17....8G}, (5) \cite{2006ApJ...637..669L}, (6) \cite{2004ApJ...615L...9W}, (7) \cite{2012MNRAS.421.1764S}, (8) \cite{2012ApJ...748...49S}, (9) \cite{2010MNRAS.402..497G}, (10) \cite{2018AJ....156...36K}, (11) \cite{2004ApJ...602..103F}, (12) \cite{2002ApJ...576...81O}, (13) \cite{2011ApJ...735...60P}, (14) \cite{2017ApJ...851...33P}, (15) \cite{2013ApJ...779..174D}, (16) \cite{2022ApJ...925..139S}, (17) \cite{2016MNRAS.463.3038X}, (18) \cite{2020AA...642A..59R}.}}
\end{deluxetable*}

\subsection{Gaussian Process Method} \label{subsec:process}

GPs are a class of statistical models, which are widely applied for modeling stochastic
processes.
For the one who is interested in the stochastic behavior of astronomical variability, 
GP provides a flexible method to model the LC with stochastic processes.
The application of GPs for astronomical time-series is discussed in a recent review \citep{2022arXiv220908940A}.
Considering a data set of $\boldsymbol{y}_n$ at coordinates $\boldsymbol{x}_n$,
the GP model consists of a mean function ${\mu}_{\theta}$($\boldsymbol{x}$) parameterized by $\boldsymbol{\theta}$ and a kernel function (covariance function) ${k}_{\alpha}(\boldsymbol{x}_{n},\boldsymbol{x}_{m})$ 
parameterized by parameters $\boldsymbol{\alpha}$ \citep{2017AJ....154..220F}.
For time-series data, the GP is one-dimensional, and the coordinates are time, $\boldsymbol{x}_n$=$t_n$.
After obtaining the likelihood function with the above information, one can use Bayesian
inference to estimate the posterior distribution over the parameter space.
 
In practical application, the key point is choosing the kernel function.
The DRW process (called Ornstein-Uhlenbeck process in physics) is widely used to describe the variability of AGNs \citep[e.g.,][]{2021Sci...373..789B}, 
and it is defined by an exponential covariance function \citep[e.g.,][]{2009ApJ...698..895K,2013ApJ...765..106Z}, 
\begin{equation}\label{eq2}
k(t_{nm})=2\sigma_{\rm DRW}^{2}\cdot \exp(-t_{nm}/\tau_{\rm DRW})\;\ ,
\end{equation}
where $t_{\rm nm} = |t_{\rm n}-t_{\rm m}|$ is the time lag between measurements $m$ and $n$.
The amplitude term ($\sigma_{\rm DRW}$) represents the amplitude of the random disturbance, and the damping (characteristic) timescale ($\tau_{\rm DRW}$) represents the timescale that the system returns to the stability after experiencing a disturbance. 
Sometimes, an excess white noise term ($\sigma^{2}_{\rm n}\delta_{\rm nm}$ where $\sigma_{\rm n}$ is the excess white noise amplitude and $\delta_{\rm nm}$ is the Kronecker $\delta$ function) is needed in the situation that there is a white noise in the LC in addition to the quoted measurement errors \citep{2018zndo...1160614F,2021Sci...373..789B}.
A more complex kernel is the stochastically driven damped simple harmonic
oscillator (SHO), which is described by a second-order differential equation \citep{2017AJ....154..220F}.
The SHO kernel has been used to model the AGN accretion disk \citep{2022ApJ...936..132Y} and jet \citep{2022ApJ...930..157Z} variability.

{\it Celerite} software package is a GP tool for a stationary process \citep{2017AJ....154..220F}.
It uses the semi-separated structure of a special covariance matrix to directly analyze and compute the GP likelihood for large data sets.
\cite{2021ApJ...907..105Y} and \citet{2021ApJ...919...58Z,2022ApJ...930..157Z} have tested the efficiency of this method for the study of AGN jet variability, 
and suggested that {\it celerite} has a strong potentiality for studying AGN variability \citep[also see][]{2021Sci...373..789B}. 
Here, we use the DRW model implemented in {\it celerite} package to model the multi-band variability of blazars. 

The Markov Chain Monte Carlo (MCMC) sampler emcee\footnote{\url{https://github.com/dfm/emcee}} is adopted to perform posterior analysis.
We assume log-uniform priors on each of the parameters.
The MCMC sampler is run for 50,000 iterations with 32 parallel walkers.
The first 20,000 steps are taken as burn-in. 
After modeling the LCs, we should estimate the fitting quality for assessing whether the fitting results are reliable, e.g.,  
whether the standardized residuals follow a Gaussian white-noise sequence.

The power spectral density (PSD) can be constructed by using the fitting results.
The DRW PSD is in the form of
\begin{equation}\label{eq3}
S(\omega)=\sqrt{\frac{8}{\pi}}\sigma_{\rm DRW}^{2}\tau_{\rm DRW}\frac{1}{1+(\omega\tau_{\rm DRW})^2}\ .
\end{equation}
It is a broken power-law form with slope 0 below the broken frequency ($f_{\rm b}$) and slope -2 above the broken frequency.
The conversion between the timescale $\tau_{\rm DRW}$ and $f_{\rm b}$ is $\tau_{\rm DRW} = 1/(2\pi f_{\rm b})$.

The LC with large cadence or insufficient length leads to a large bias on the characteristic timescale derived from modeling.
If the timescale is larger than the mean cadence of LC and less than 1/10 of the length of LC, 
the measurement of the damping timescale from the LC is reliable \citep{2021Sci...373..789B}.

\section{Results} \label{sec:results}
\subsection{Results of 3C 273, PKS 1510-089 and BL Lac} \label{subsec:individual}

We first analyze the multi-band variability of the three individual sources, 3C 273, PKS 1510-089, and BL Lac.
We present the {\it celerite} fitting results of the LC for each source in the following.
The measured timescales given in the main text are with errors in 95$\%$ confidence intervals.

\begin{figure}
    \centering
    {\includegraphics[width=16cm]{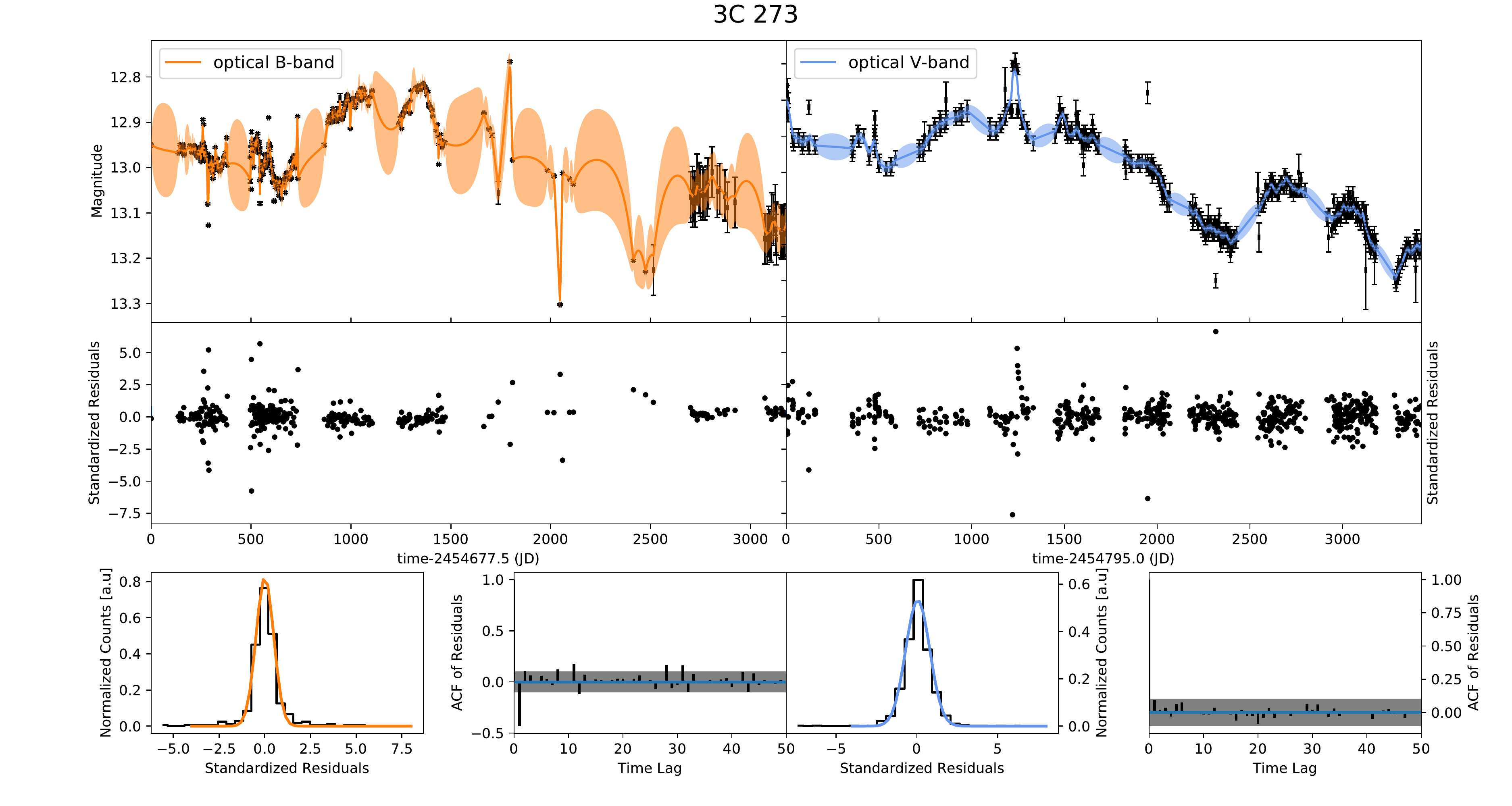}}
    \caption{DRW fitting results of 3C 273 in $B$-band (left panel) and $V$-band (right panel). 
    For each column, the top panel presents the observed LC (black points) and the modeled LC (orange/blue line). 
    We show the standardized residuals (black points) in the middle panel.
    In the bottom panel, there are two parts. The probability density of standardized residuals (black ladder diagram) as well as the best-fit normal distribution (orange/blue solid line) are shown in the left part.  
    The ACF of residuals with the 95$\%$ confidence limits of the white noise (the gray region) are shown in the right part. 
\label{fig:3C 273 fit}}
\end{figure}

\begin{figure}
    \centering
    {\includegraphics[width=8cm]{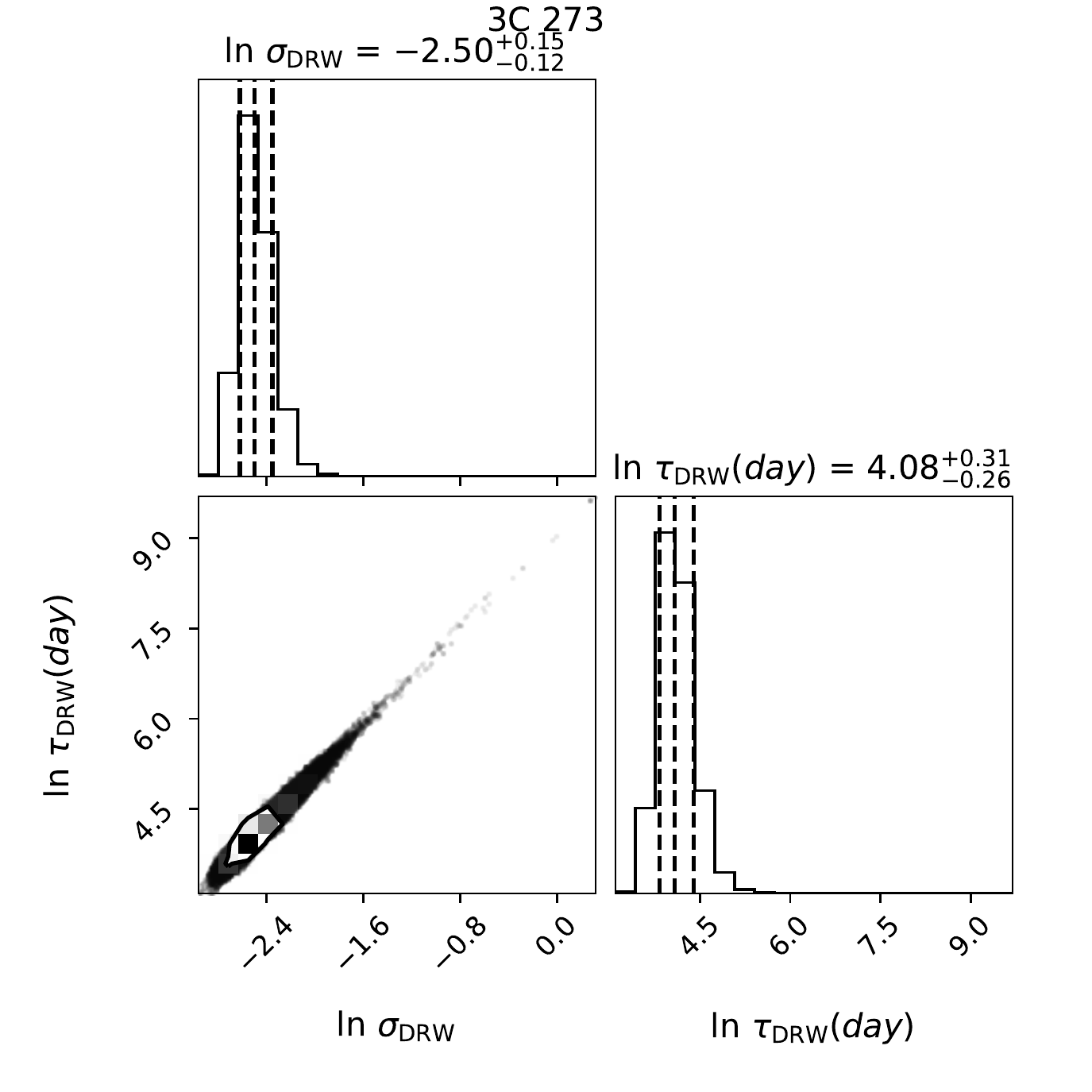}}
    {\includegraphics[width=8cm]{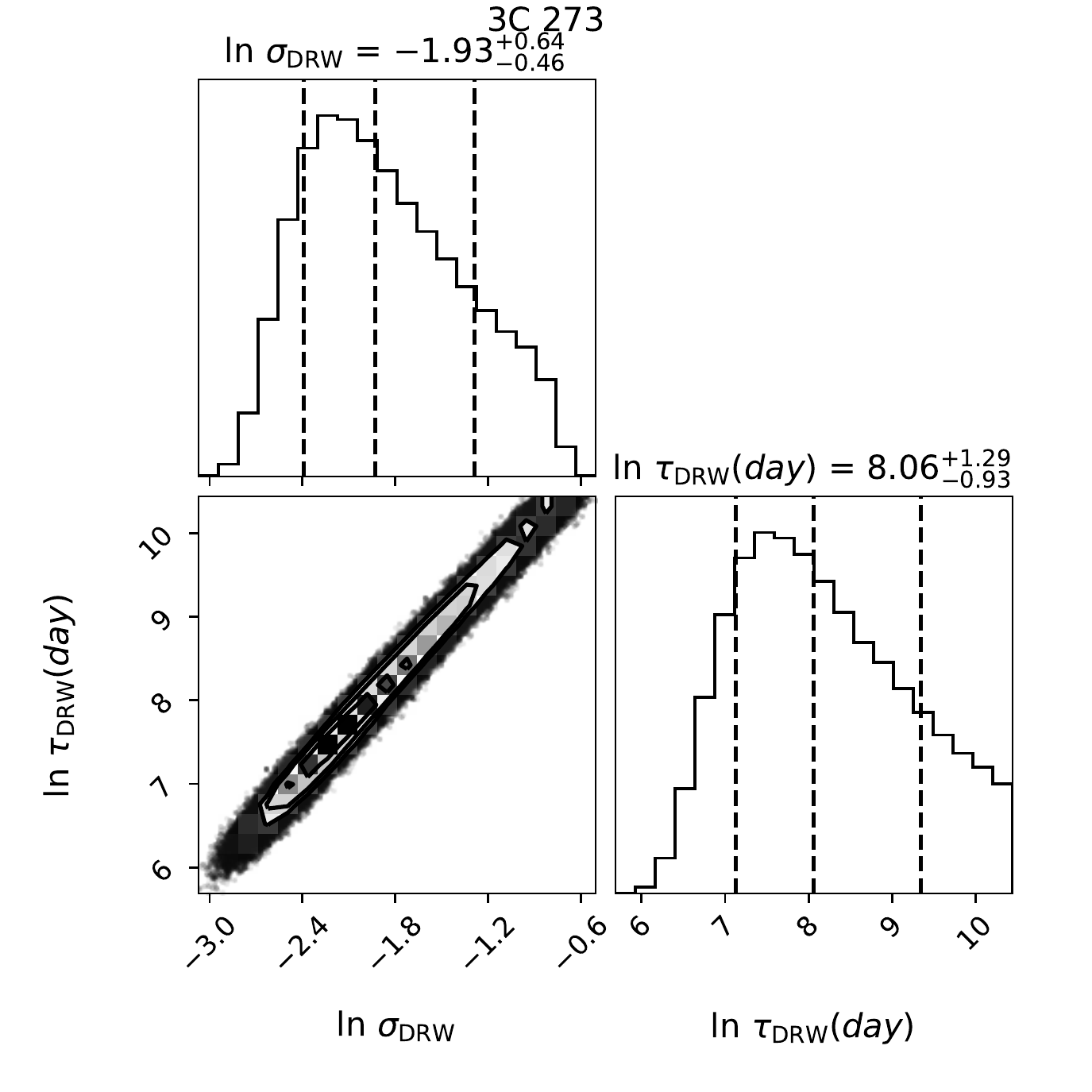}}
    \caption{Posterior probability densities of model parameters for 3C 273 in $B$-band (left) and $V$-band (right). 
                The vertical dotted lines mark the median value and 68$\%$ confidence intervals of the distribution of the parameter. 
\label{fig:3C 273 param}}
\end{figure}

\begin{figure}
    \centering
    {\includegraphics[width=8cm]{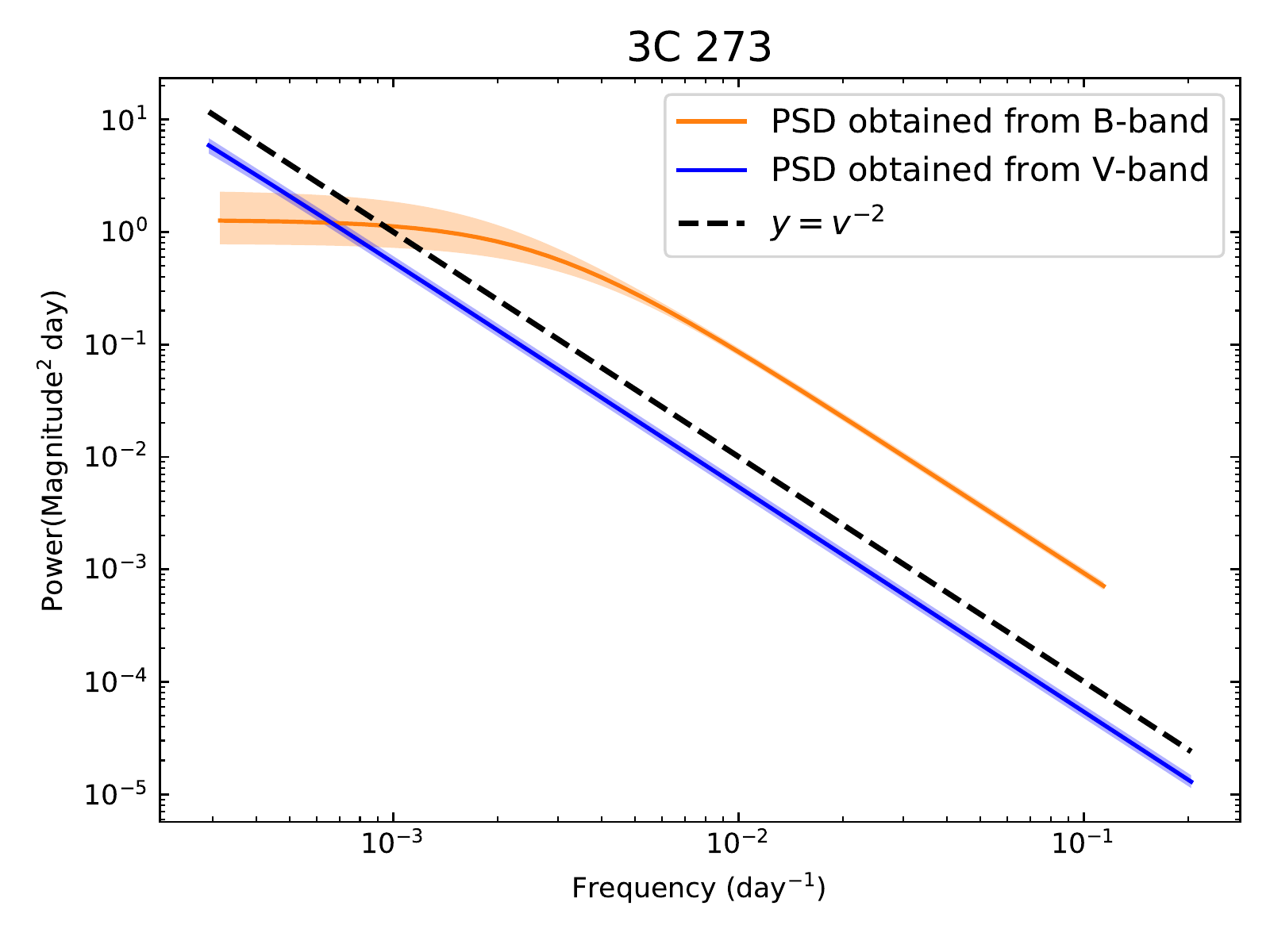}}
    \caption{$B$-band and $V$-band PSDs of 3C 273 constructed from the modeling results with DRW model. 
    The orange line is $B$-band PSD, and the blue line is $V$-band PSD. The corresponding color region denotes the 1$\sigma$ confidence interval. 
    The dashed black line is a reference line with a slope of -2.
\label{fig:3C 273 psd}}
\end{figure}

\begin{figure}
    \centering
    {\includegraphics[width=16cm]{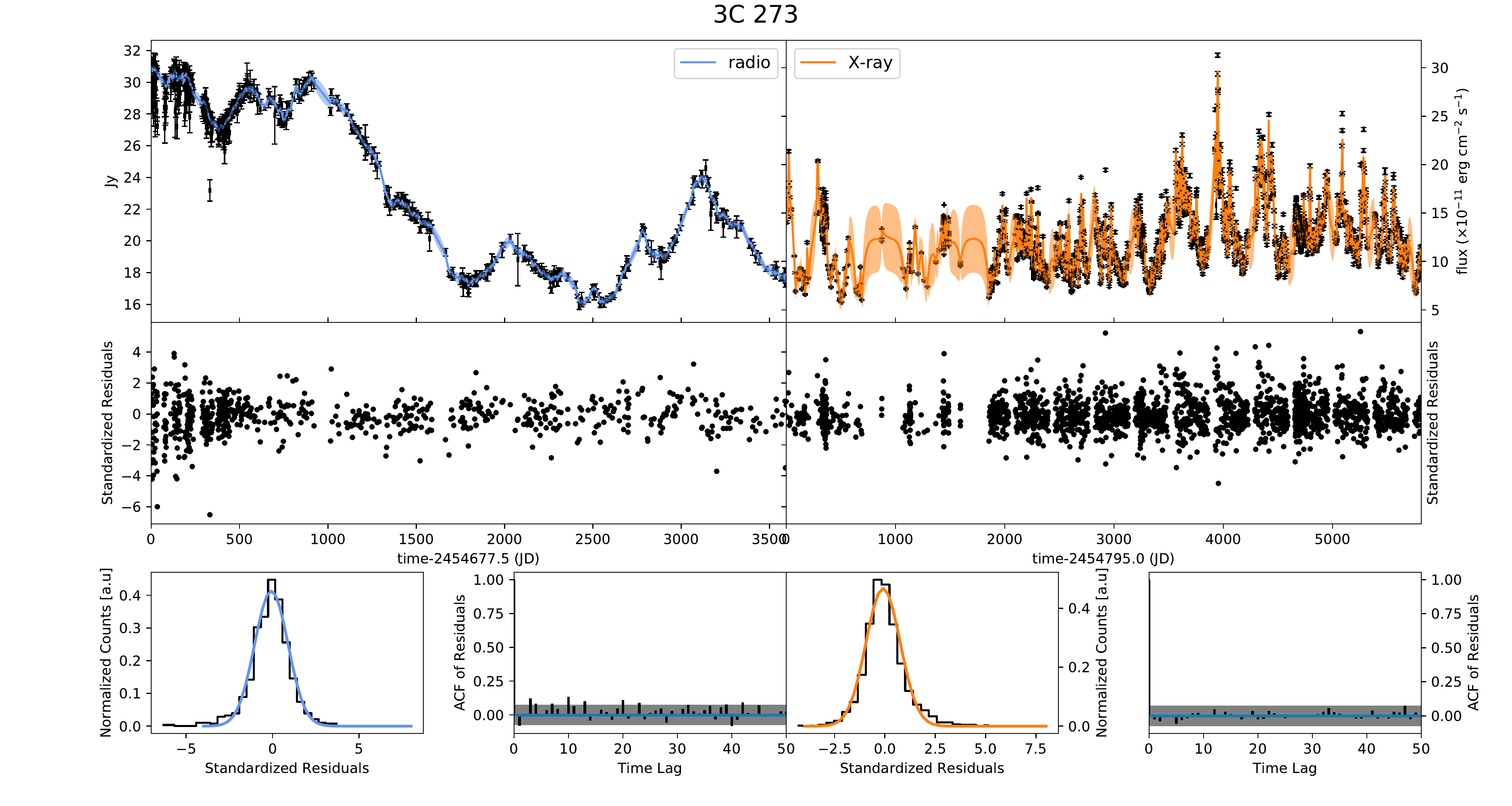}}
    \caption{DRW fitting results of 3C 273 for radio (left panel) and X-ray (right panel) data. 
    The symbols and lines are the same as those in Figure~\ref{fig:3C 273 fit}.
\label{fig:3C 273 X+radio}}
\end{figure}

\begin{figure}
    \centering
    {\includegraphics[width=8cm]{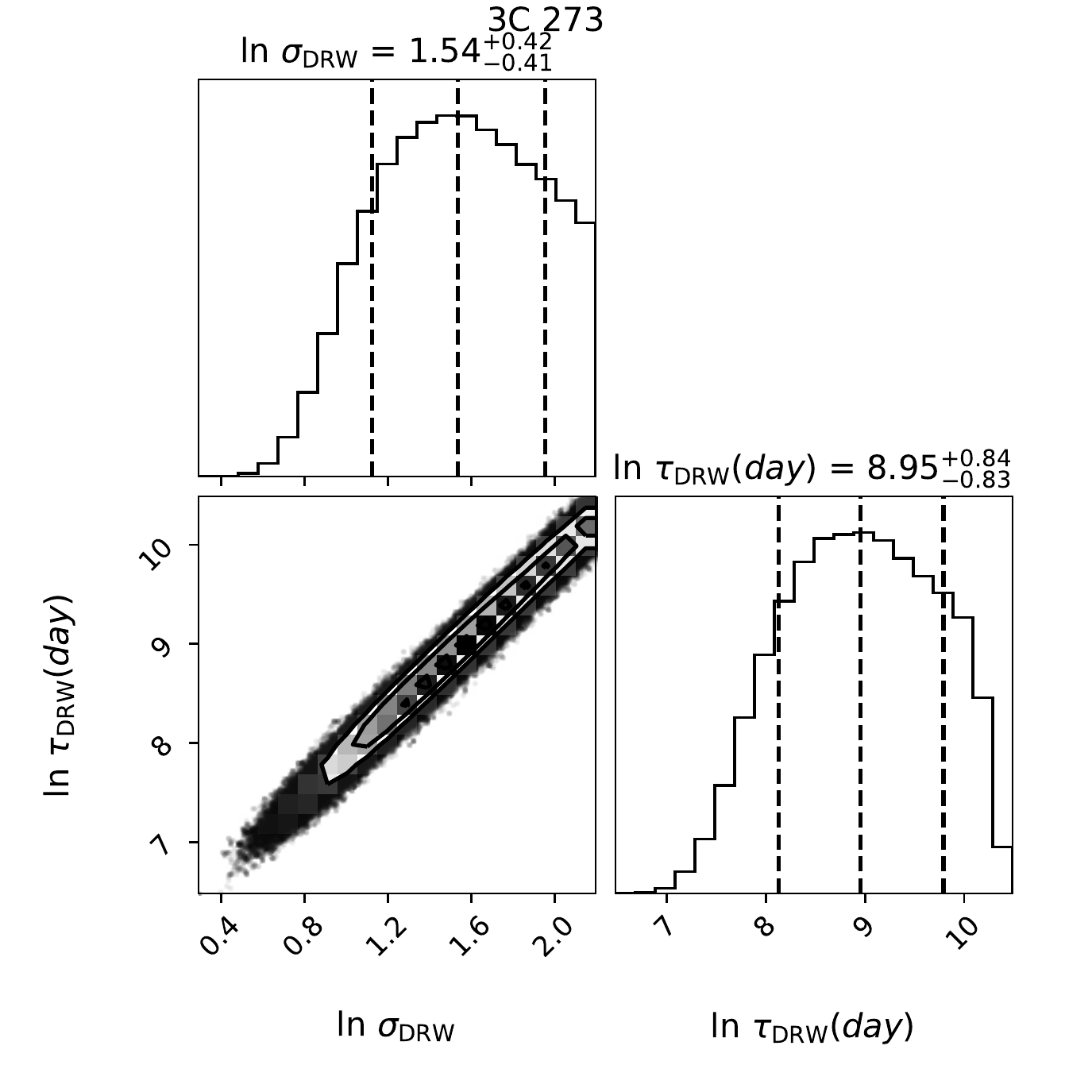}}
    {\includegraphics[width=8cm]{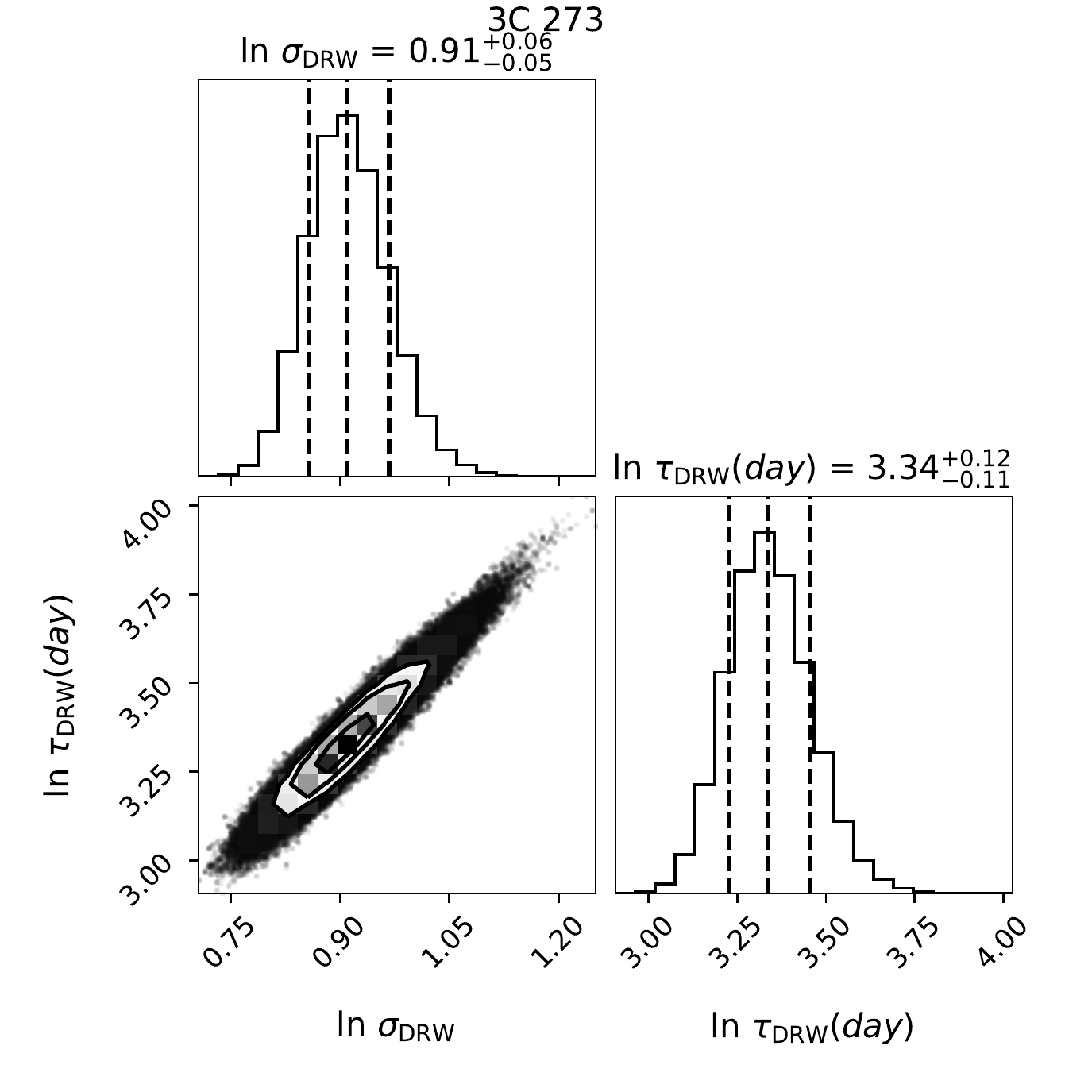}}
    \caption{Posterior probability densities of model parameters for 3C 273 in radio (left) and X-ray (right) energies. 
          The vertical dotted lines mark the median value and 68$\%$ confidence intervals of the distribution of the parameter. 
\label{fig:X+radio params 3C273}}
\end{figure}

\begin{figure}
    \centering
    {\includegraphics[width=8cm]{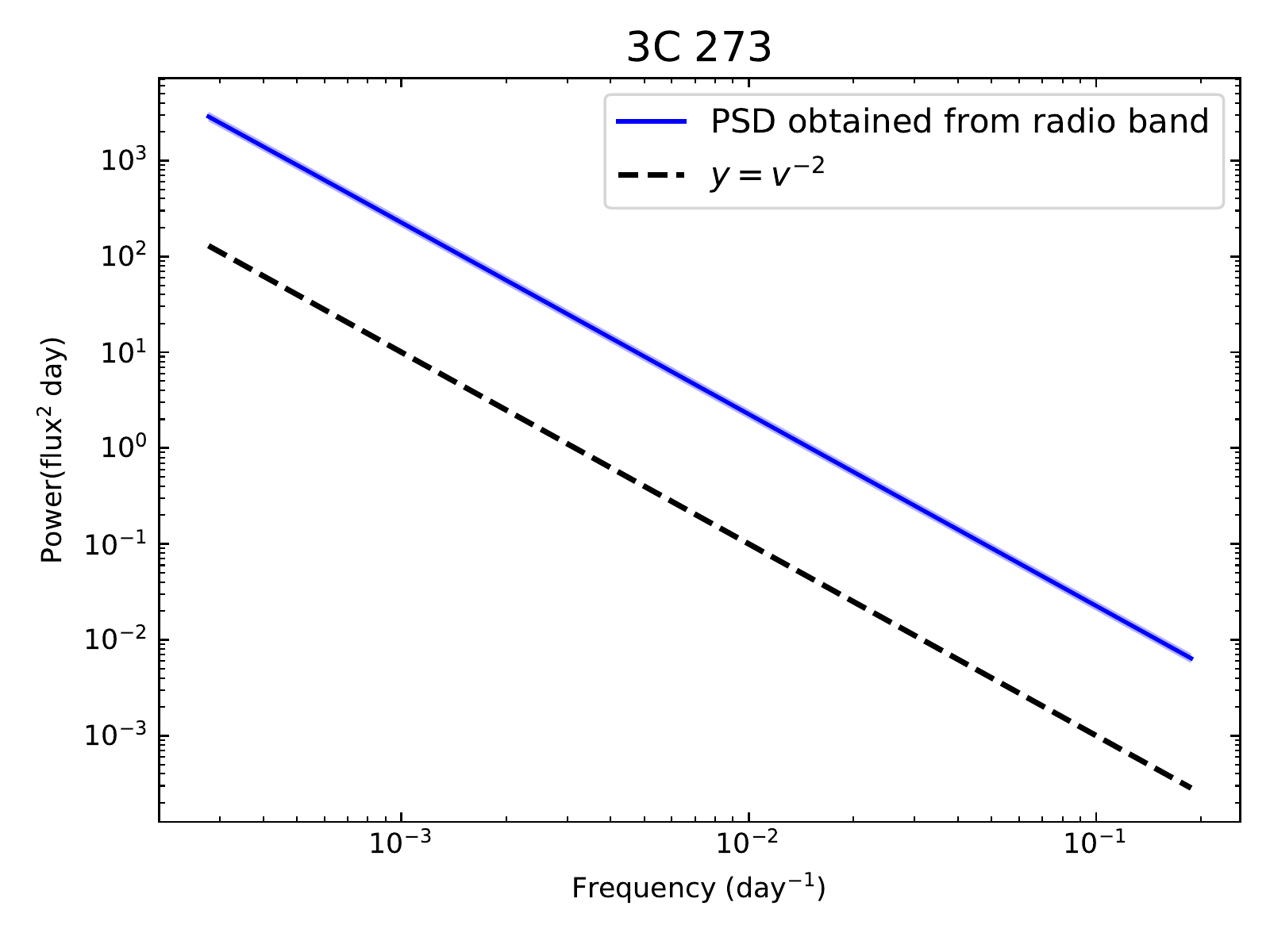}}
    {\includegraphics[width=8cm]{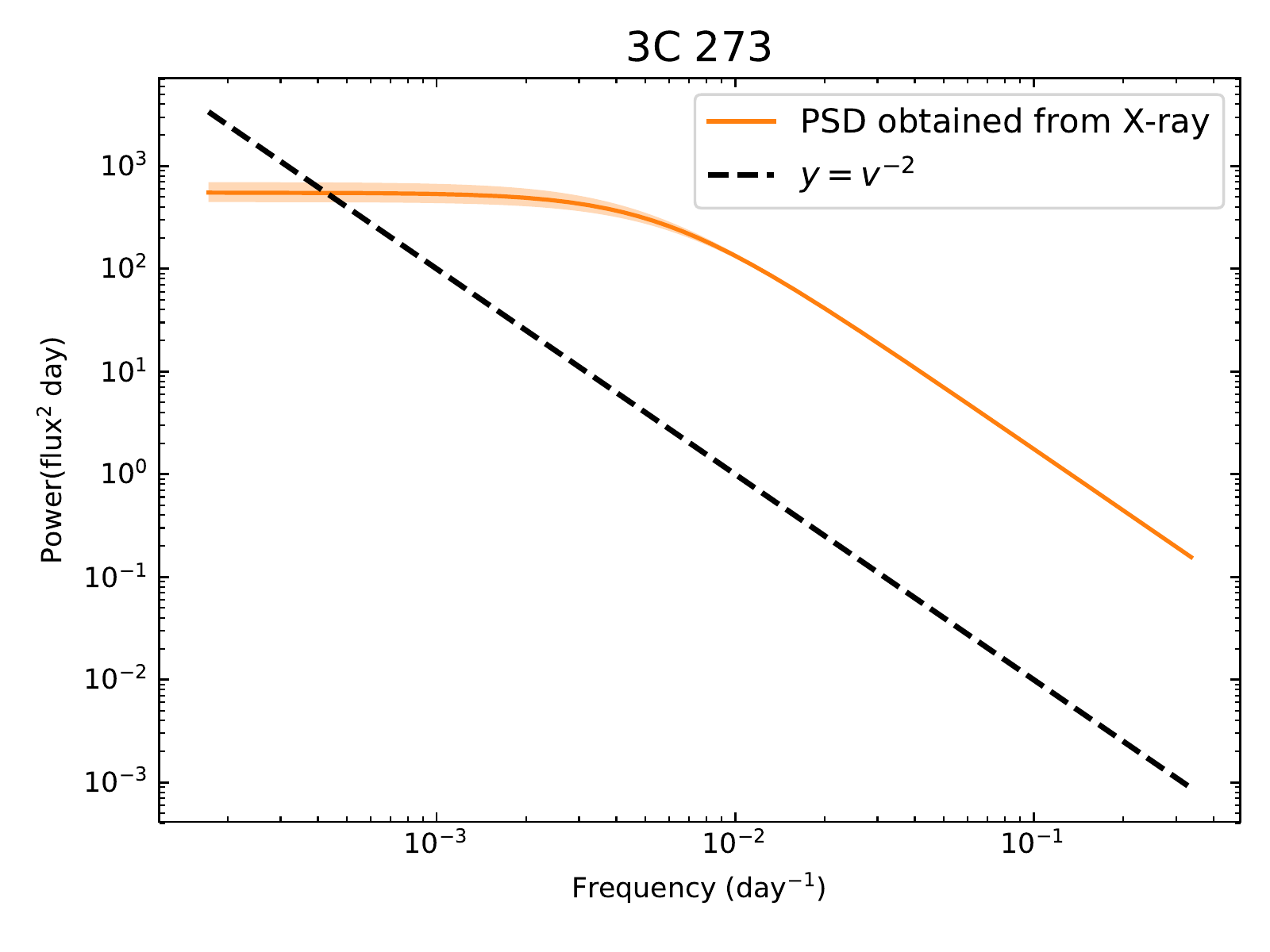}}
    \caption{Radio and X-ray PSDs constructed from modeling LCs of 3C 273 with DRW model. 
     The symbols and lines are the same as those in Figure~\ref{fig:3C 273 psd}. 
\label{fig:X+radio psd 3C273}}
\end{figure}

For 3C 273, the optical data in both $B$ and $V$ bands are available. 
We show the modeling results in Figure~\ref{fig:3C 273 fit}, in which the left panel is for $B$-band LC and the right is for $V$-band LC.
The DRW model can agree well with both LCs.
Looking at the distribution of standardized residuals and the auto-correlation function (ACF) of standardized residuals \citep[see details in][]{2022ApJ...930..157Z},
we believe the characteristic of each LC has been captured successfully. 
Through MCMC sampling, we get the posterior probability density distributions of two parameters ($\sigma_{\rm DRW}$ and $\tau_{\rm DRW}$) and show them in Figure~\ref{fig:3C 273 param}.
The values are listed in Table~\ref{tab:fit}.
The results are different between the two bands.
The parameters can be constrained by the $B$-band data but with large uncertainties, e.g., $\tau_{\rm DRW}=59^{+41}_{-28}$ days. 
Comparing the timescale with the cadence and the length of the LC, we believe that the $B$-band timescale is reliable.
A broken frequency corresponding to the characteristic timescale is shown in the $B$-band PSD (Figure~\ref{fig:3C 273 psd}).
While the $V$-band timescale is $\approx3200$ days, very close to the length of the LC.
This means that the $V$-band timescale is unreliable, which is also confirmed by the single power-law PSD (Figure~\ref{fig:3C 273 psd}).
We show the modeling results of X-ray LC, the posterior probability density distribution of parameters, 
and the PSD in the right panel in Figure~\ref{fig:3C 273 X+radio}, Figure~\ref{fig:X+radio params 3C273} and Figure~\ref{fig:X+radio psd 3C273} respectively.
The values of the parameters can be found in Table~\ref{tab:x-ray fit}.
It is shown that the DRW model can describe the X-ray variability of 3C 273. 
The parameters are well constrained.
The X-ray PSD presents a broken frequency that corresponds to a timescale of $\tau_{\rm DRW}=28^{+7}_{-6}$ days.
We give the radio results together with the X-ray results.
The radio PSD (the left panel of Figure~\ref{fig:X+radio psd 3C273}) of 3C 273 is a single power law.
The radio timescale is too large to be reliable.

\begin{figure}
    \centering
    {\includegraphics[width=16cm]{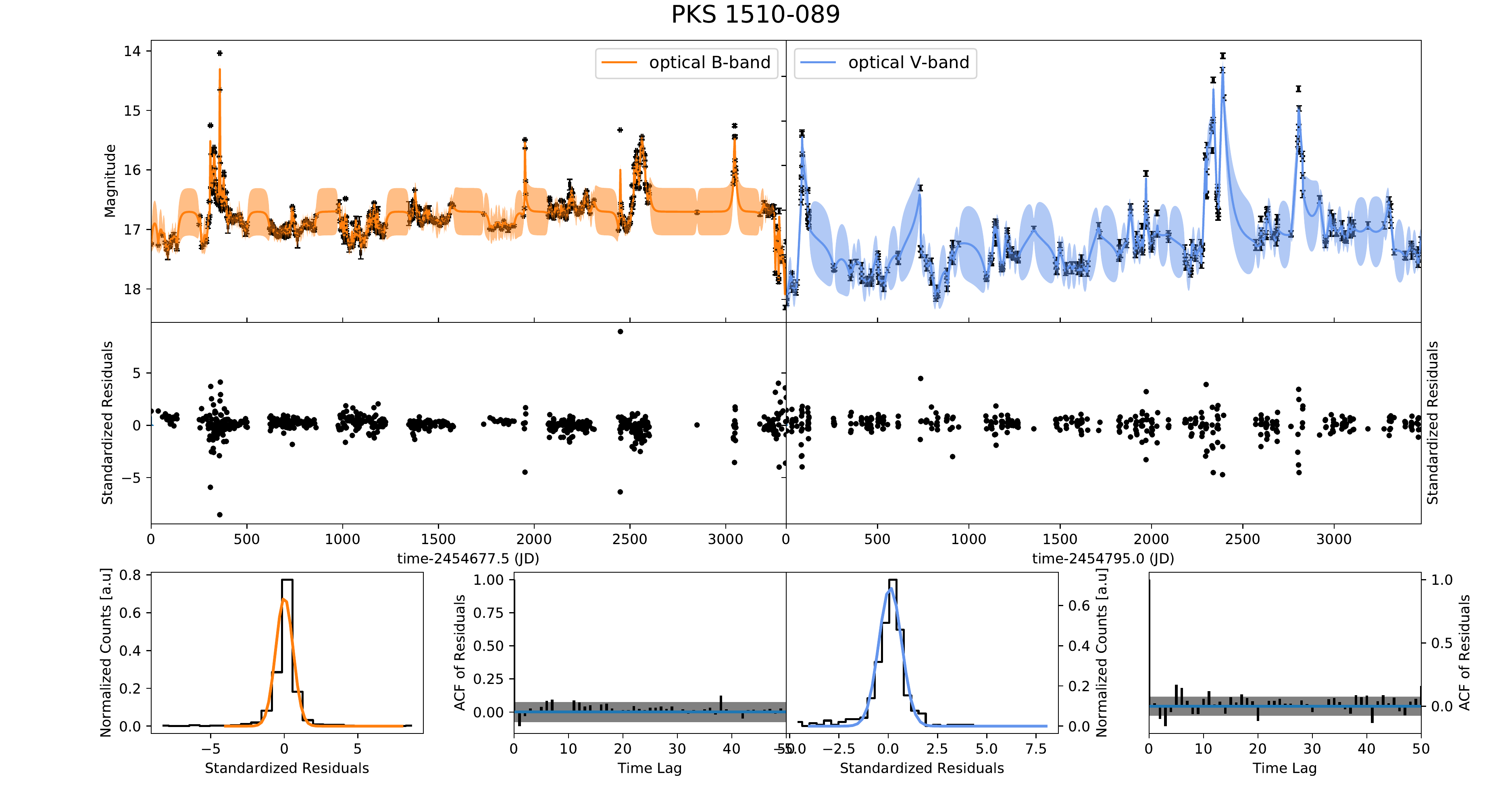}}
    \caption{DRW fitting results of $B$-band (left panel) and $V$-band (right panel) LCs for PKS 1510-089. 
    The symbols and lines are the same as those in Figure~\ref{fig:3C 273 fit}.
\label{fig:PKS1510-089 fit}}
\end{figure}

\begin{figure}
    \centering
    {\includegraphics[width=8cm]{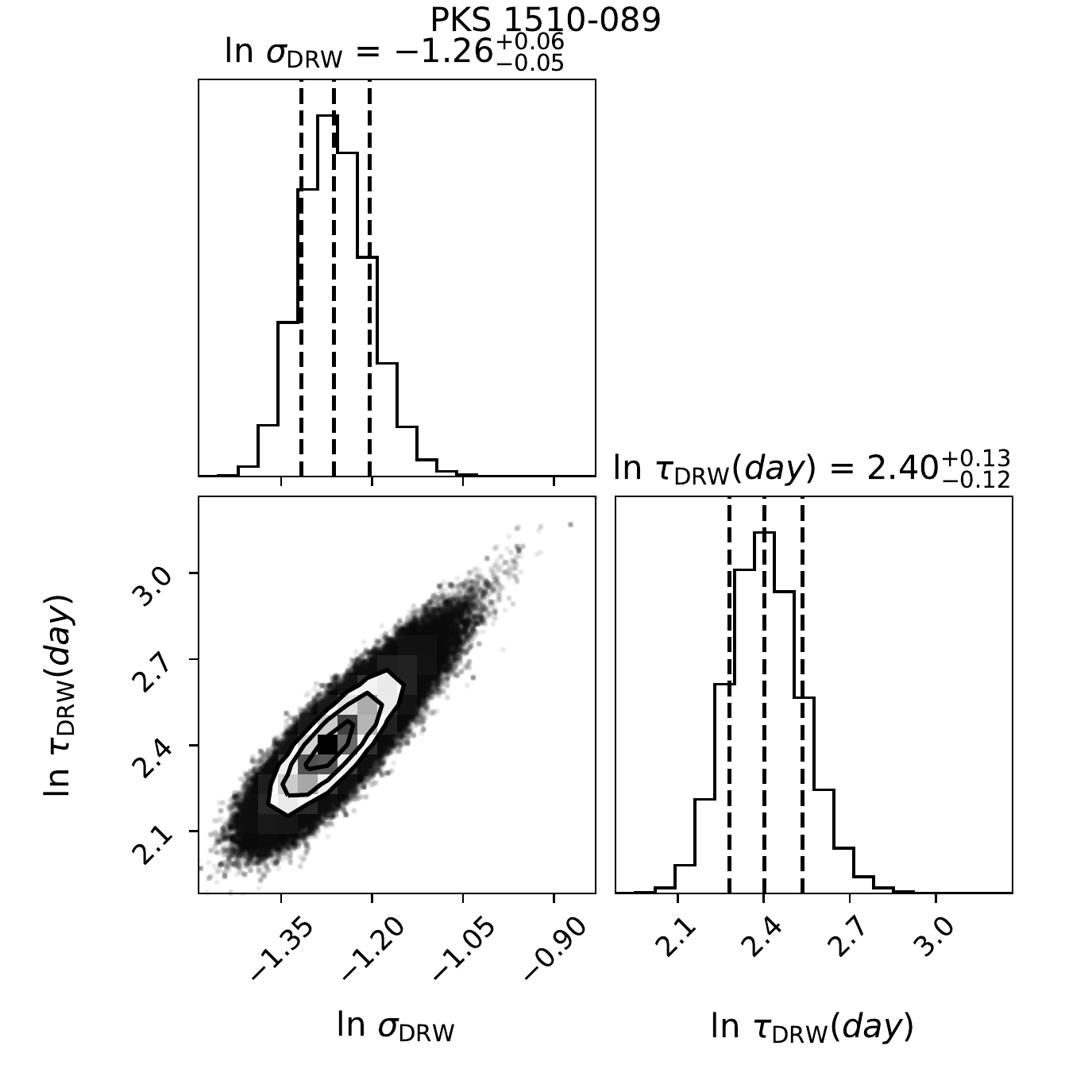}}
    {\includegraphics[width=8cm]{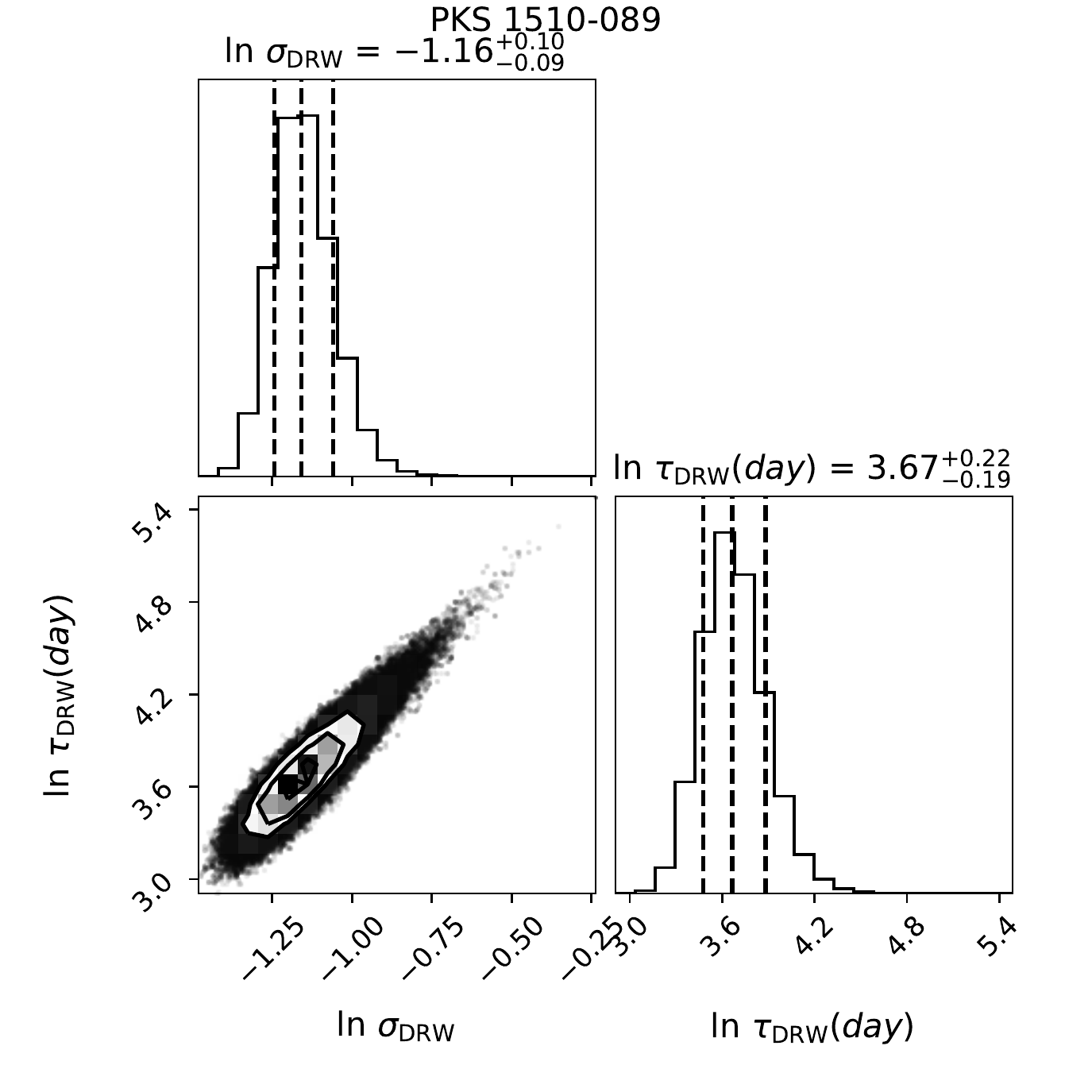}}
    \caption{Posterior probability densities of model parameters of $B$-band (left) and $V$-band (right) LCs for PKS 1510-089. The symbols and lines are the same as those in Figure~\ref{fig:3C 273 param}. 
\label{fig:PKS1510-089 param}}
\end{figure}

\begin{figure}
    \centering
    {\includegraphics[width=8cm]{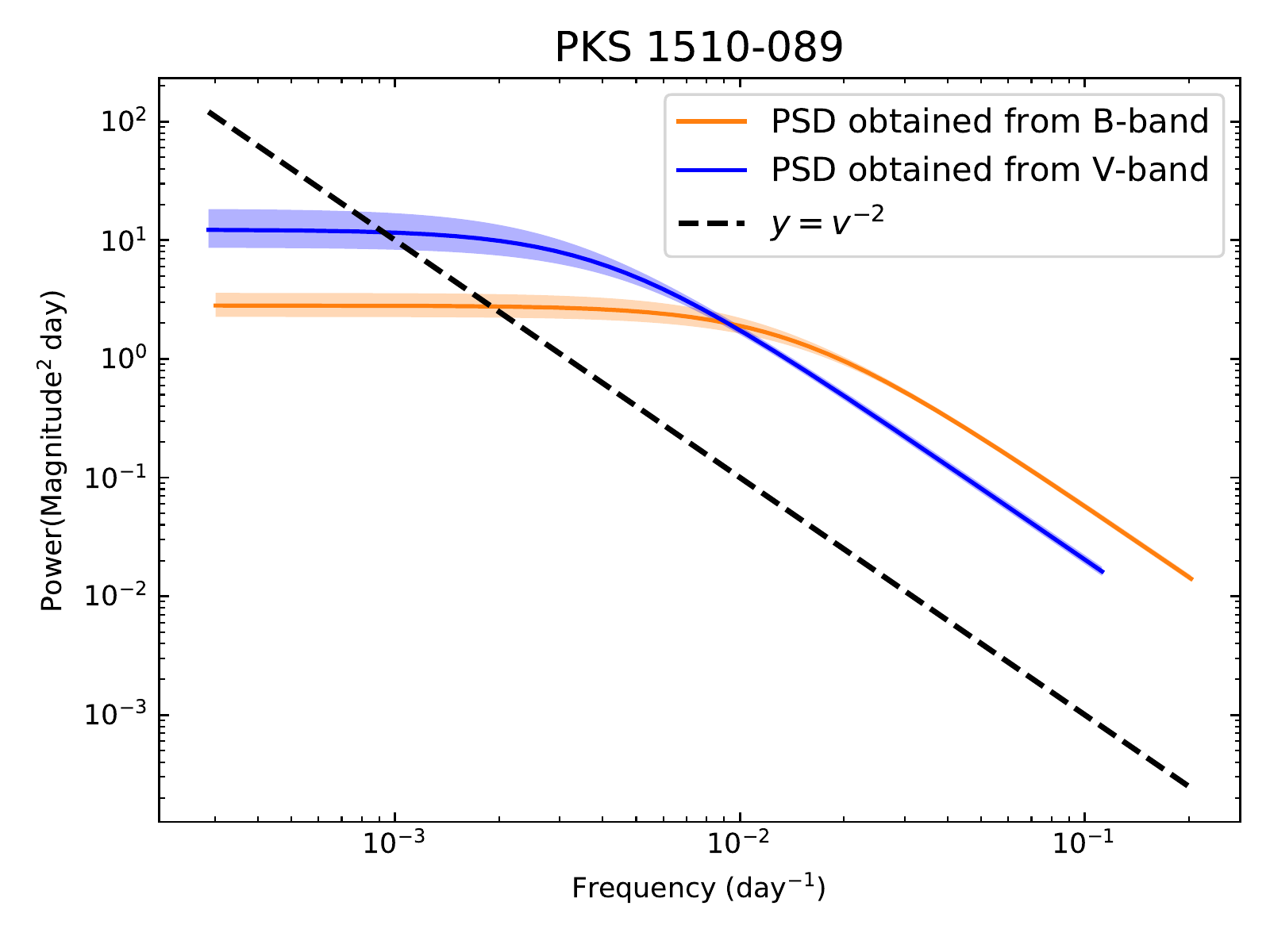}}
    \caption{$B$ and $V$-band PSDs of PKS 1510-089 constructed from the modeling results with the DRW model. The symbols and lines are the same as those in Figure~\ref{fig:3C 273 psd}. 
\label{fig:PKS1510-089 psd}}
\end{figure}

\begin{figure}
    \centering
    {\includegraphics[width=5.5cm]{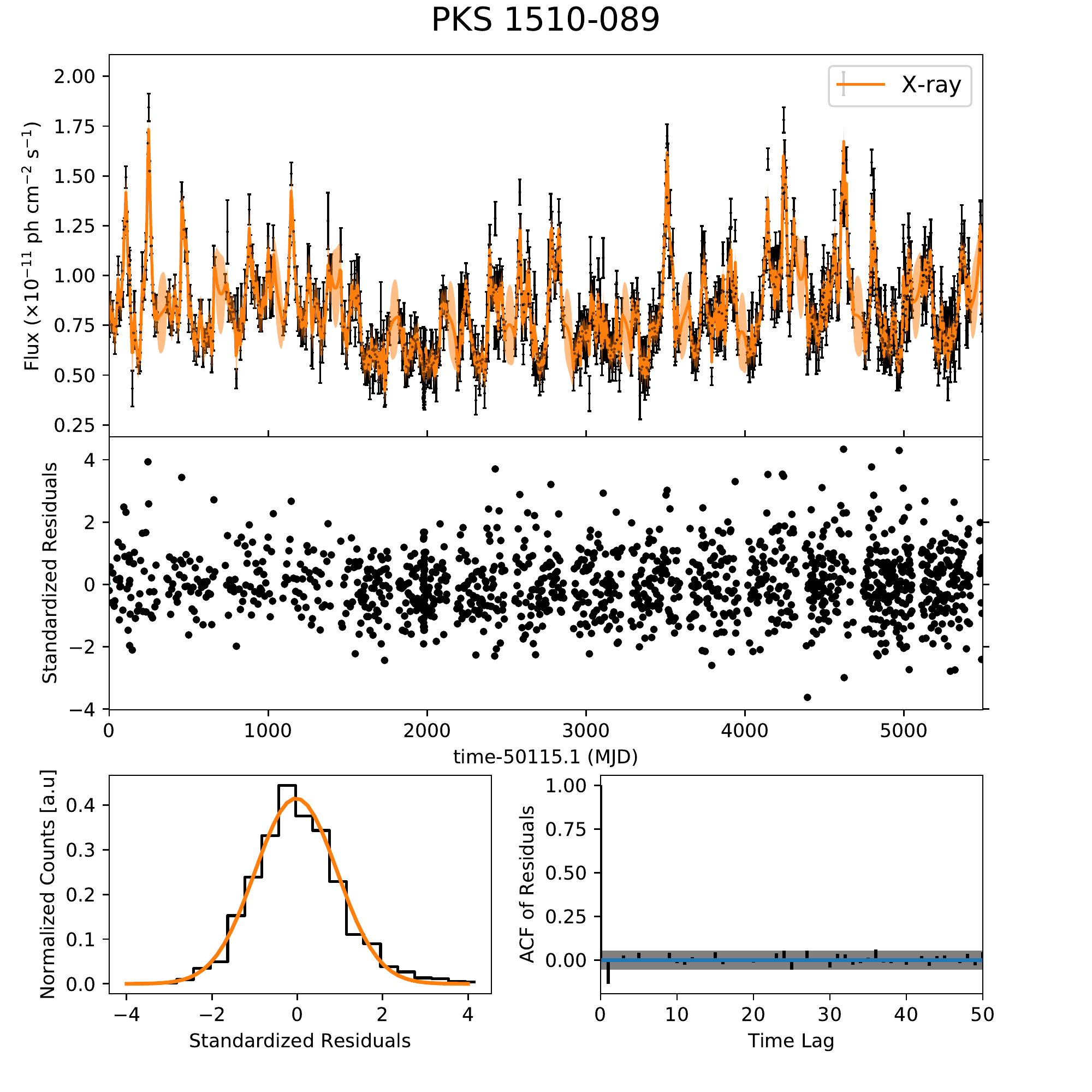}}
    {\includegraphics[width=5.5cm]{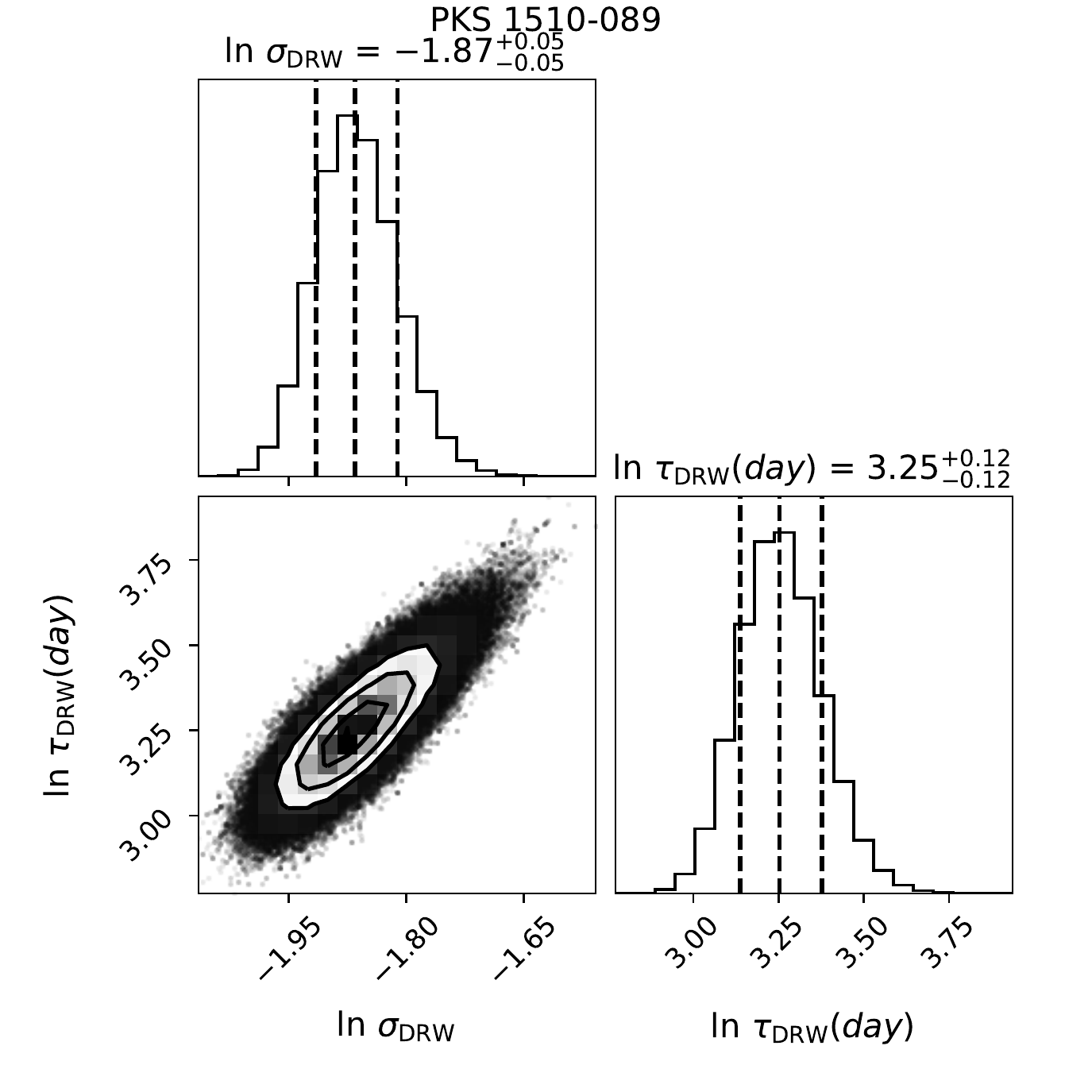}}
    {\includegraphics[width=5.5cm]{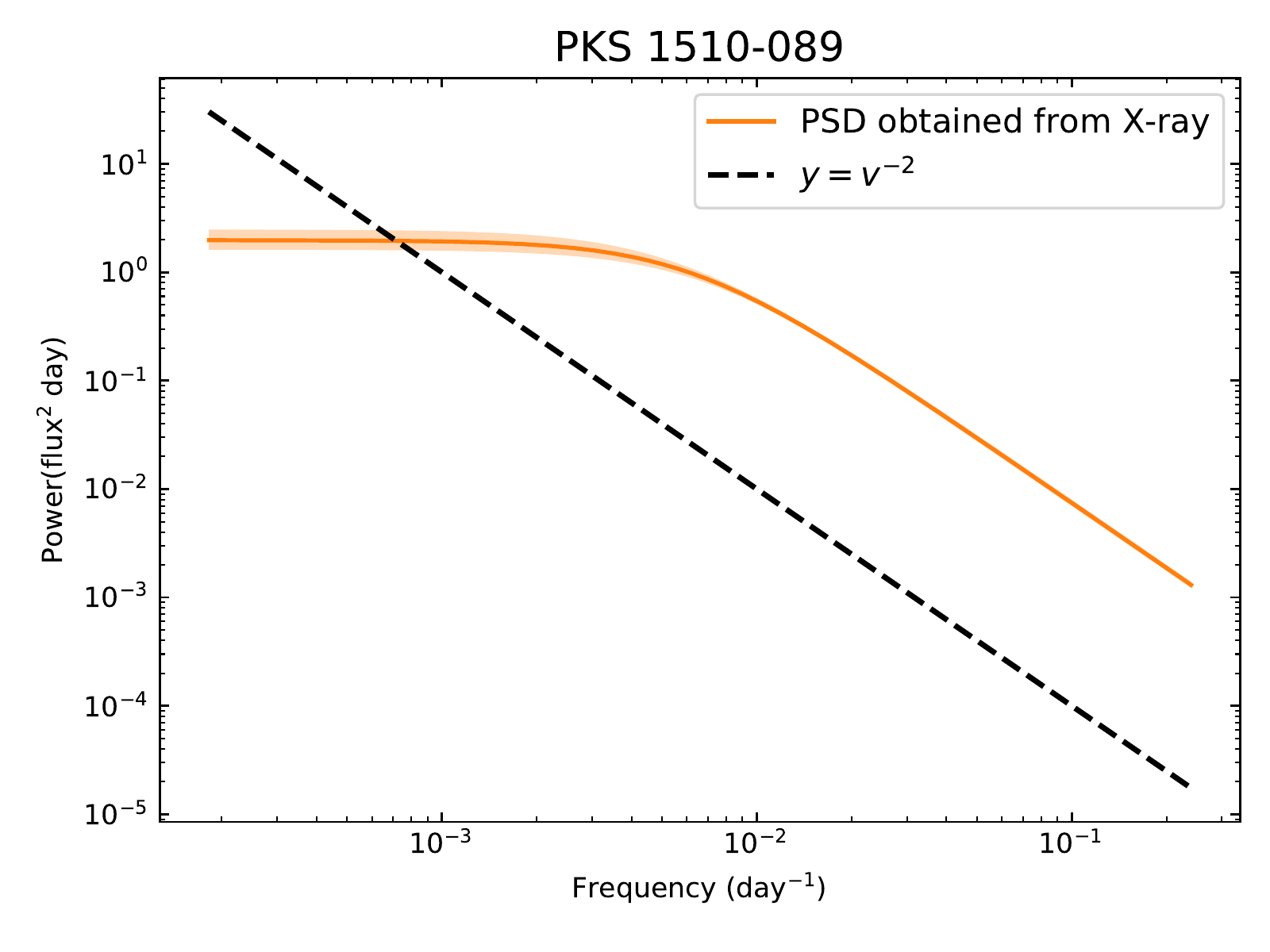}}
    \caption{Modeling results of the X-ray LC (left), the posterior probability density distribution of parameters (middle), and the X-ray PSD (right) for PKS 1510-089. 
\label{fig:X-ray PKS1510-089}}
\end{figure}

For PKS 1510-089, the $V$ and $B$-band LCs can be  
described by the DRW model (Figure~\ref{fig:PKS1510-089 fit} and Figure~\ref{fig:PKS1510-089 param}).
The $V$-band $\tau_{\rm DRW}$ of $39^{+18}_{-14}$ days is larger than the $B$-band $\tau_{\rm DRW}$ of $11\pm 3$ days (Table~\ref{tab:X-optical-gamma tau}).
As expected, we get a smaller value of $f_{\rm b}$ in $V$-band PSD (Figure~\ref{fig:PKS1510-089 psd}).
The X-ray LC of PKS 1510-089 also can be fitted well by the DRW model (Figure~\ref{fig:X-ray PKS1510-089}).
The parameters are well 
constrained (Table~\ref{tab:x-ray fit}), and the PSD is in the form of typical DRW PSD.
A trusted timescale $\tau_{\rm DRW}=26\pm 3$ days is obtained.

\begin{figure}
    \centering
    {\includegraphics[width=16cm]{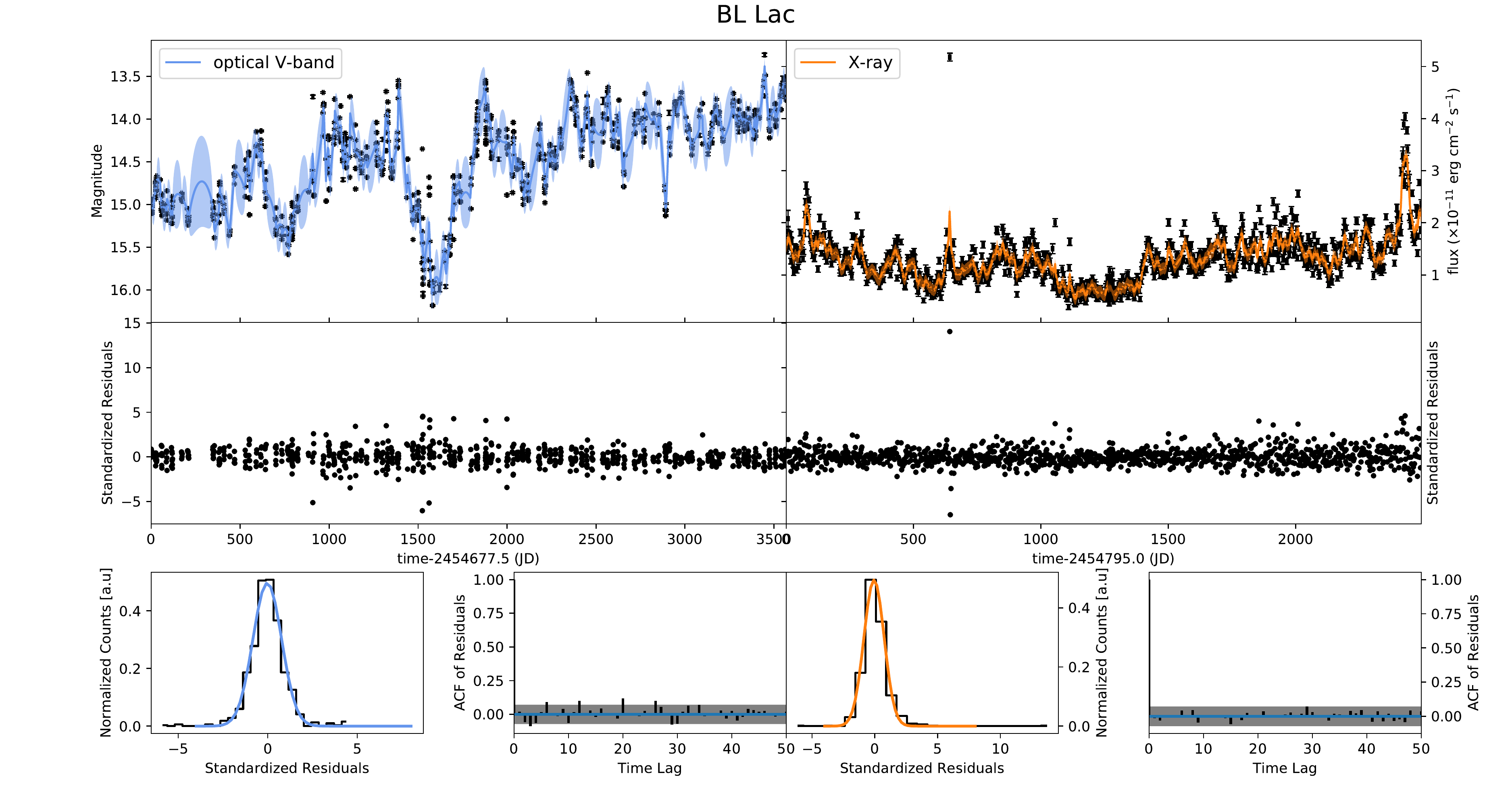}}
    \caption{DRW fitting results of $V$-band (left panel) and X-ray data (right panel) for BL Lac. 
    The symbols and lines are the same as those in Figure~\ref{fig:3C 273 fit}.
\label{fig:BLLac fit}}
\end{figure}

\begin{figure}
    \centering
    {\includegraphics[width=8cm]{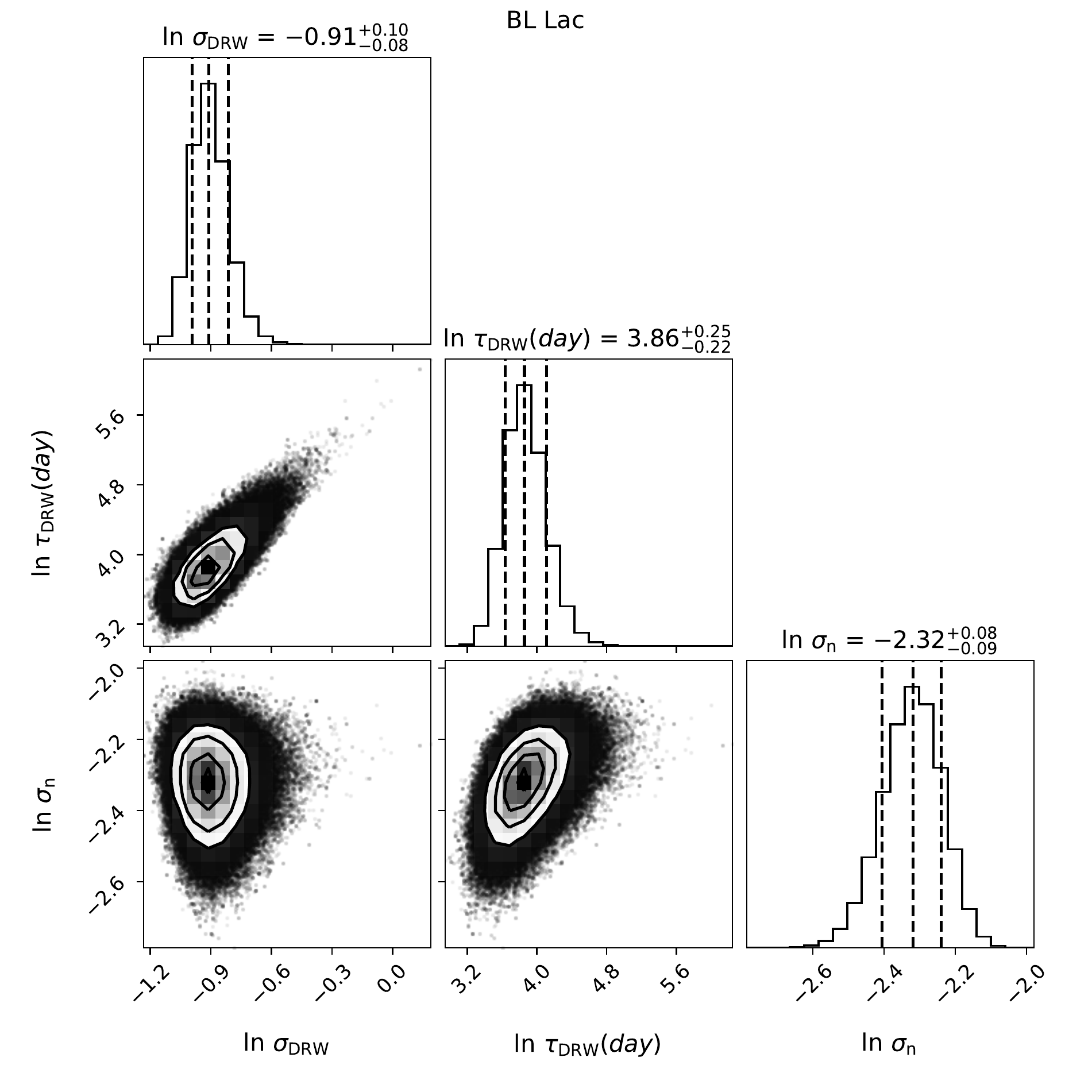}}
    {\includegraphics[width=8cm]{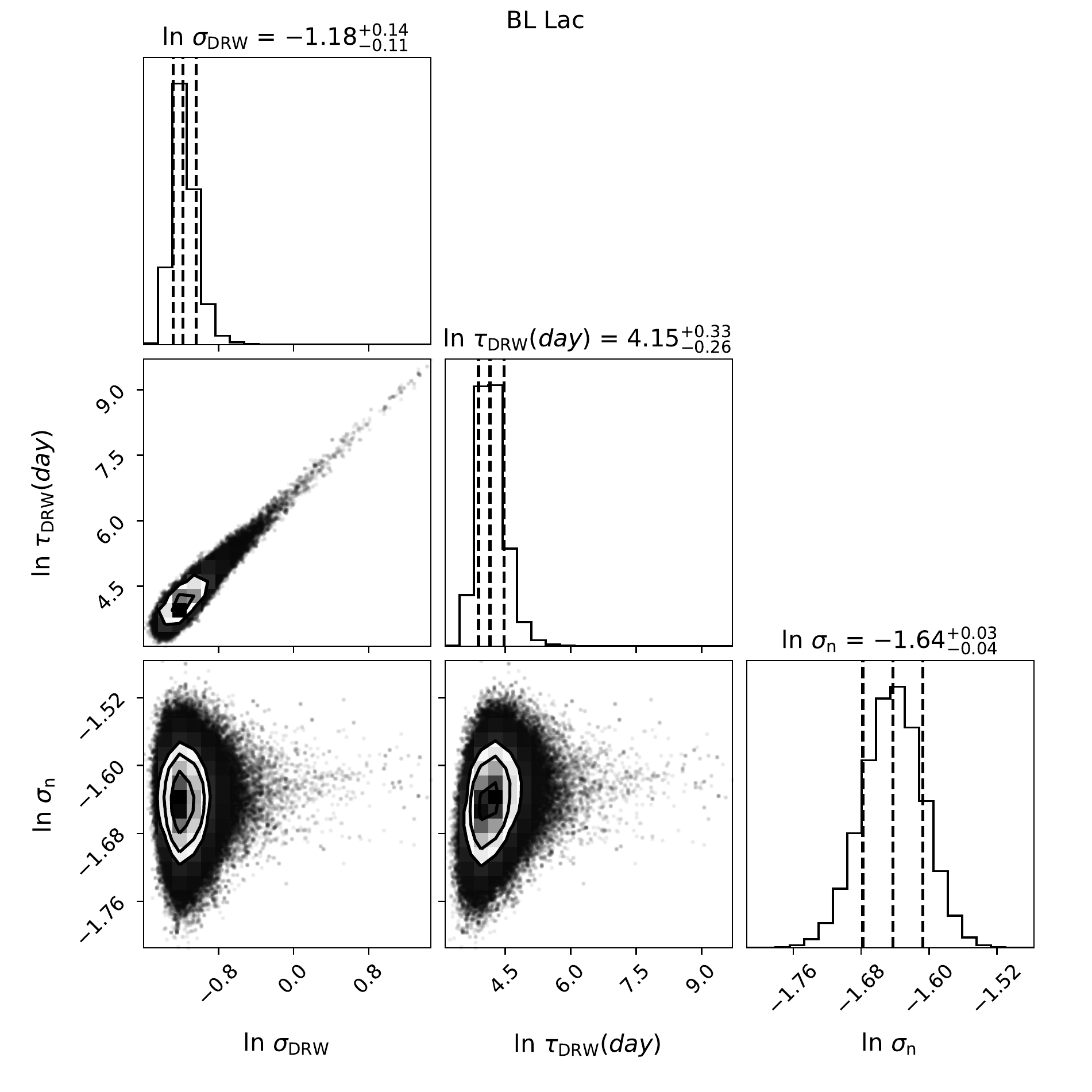}}
    \caption{$V$-band (left) and X-ray (right) posterior probability densities of model parameters for BL Lac. The symbols and lines are the same as those in Figure~\ref{fig:3C 273 param}. 
\label{fig:BLLac param}}
\end{figure}

\begin{figure}
    \centering
    {\includegraphics[width=8cm]{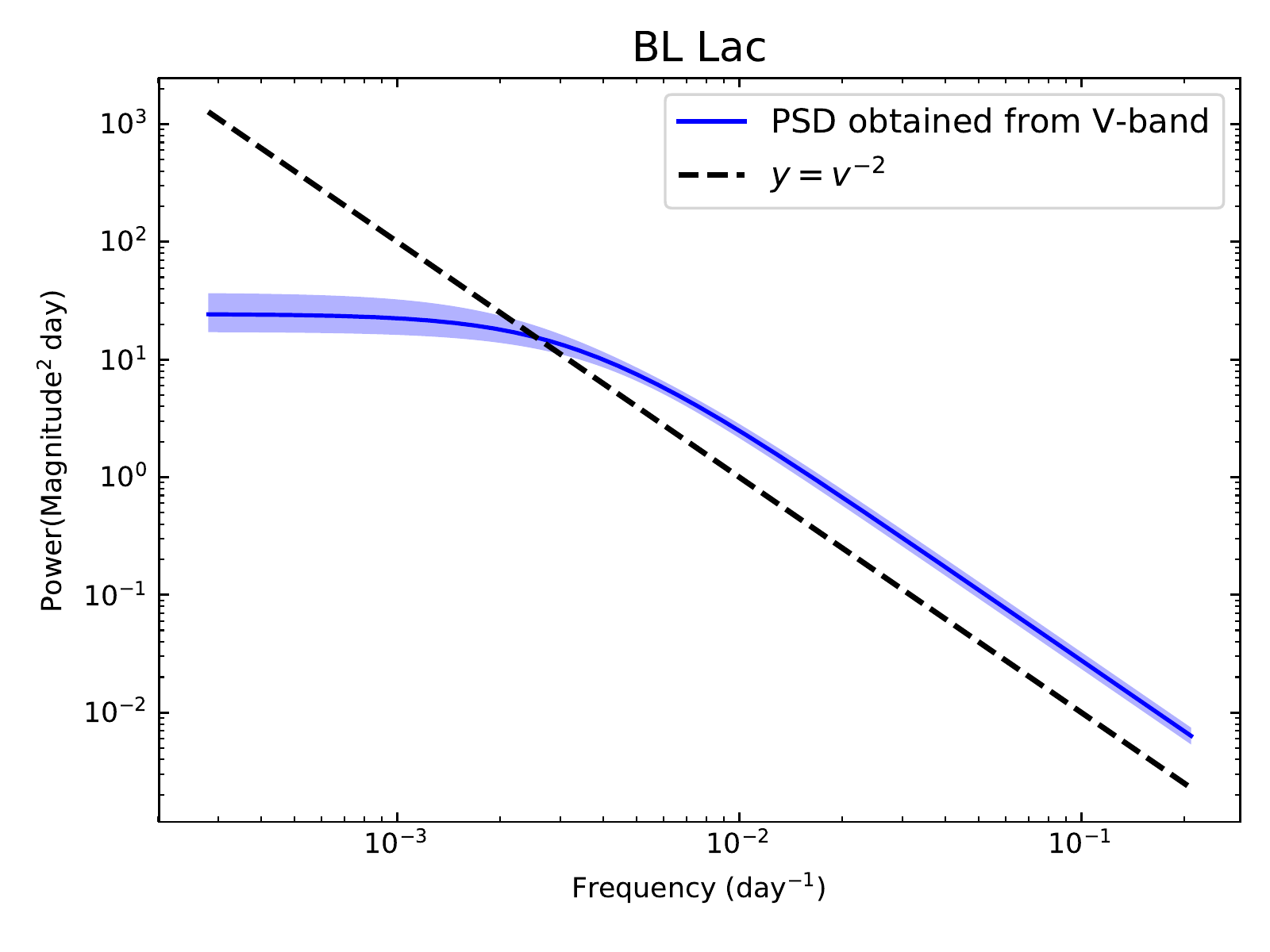}}
    {\includegraphics[width=8cm]{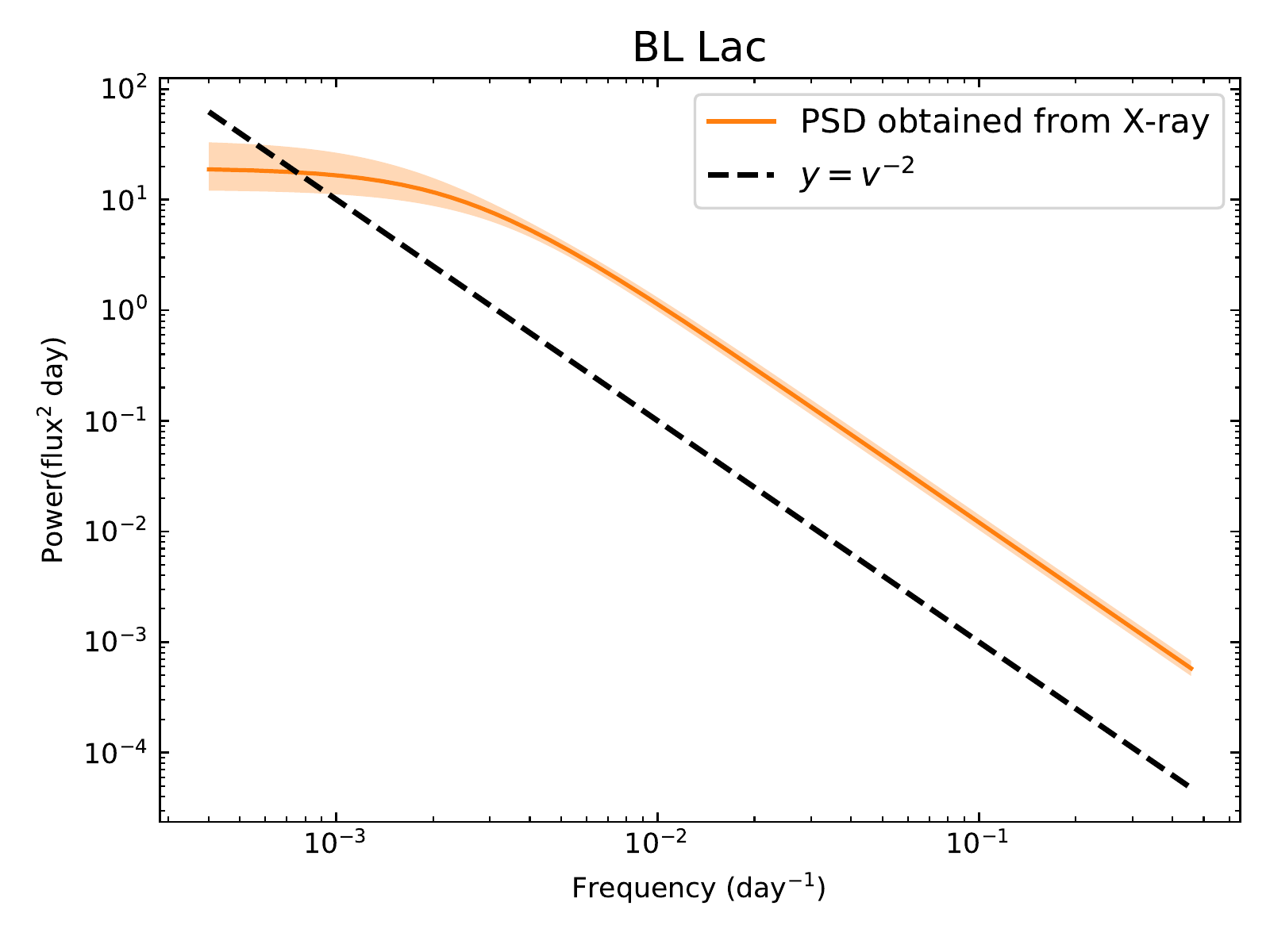}}
    \caption{$V$-band (left) and X-ray (right) PSDs for BL Lac. The symbols and lines are the same as those in Figure~\ref{fig:3C 273 psd}. 
\label{fig:BLLac psd}}
\end{figure}

For BL Lac, only the $V$-band and X-ray data are available.
For the X-ray data, there are two large gaps in the first 2800 days of LC, we then take the following 2500 days of LC for analysis.
When modeling the two LCs of BL Lac, we get poor fitting (ACF of residuals deviating from the white noise) with the two-parameter DRW model.
An excess white noise term is then added to the DRW model, and we model the LCs with the three-parameter DRW model again.
The modeling of LC, the posterior distribution of parameters, and the broken power-law PSDs are shown in Figure~\ref{fig:BLLac fit}, Figure~\ref{fig:BLLac param}, and Figure~\ref{fig:BLLac psd}, respectively.
Optical results are shown in the left panels and the X-ray results are shown in the right panels.
One can see that the three-parameter DRW can fit the LCs well.
Note that the highest flux point in the LC is poorly fitted.
After removing the highest flux point, the modeling results are unchanged.
We obtain the X-ray timescale of $\tau_{\rm DRW}=63^{+49}_{-30}$ days and $V$-band $\tau_{\rm DRW}$ of $47^{+26}_{-19}$ days (Table~\ref{tab:X-optical-gamma tau}).

We applied the SHO model to the optical and X-ray data of BL Lac. 
The fitting is not improved significantly, and the parameters cannot be constrained.
This suggests that the SHO model is not a good choice.
The DRW including an additional white noise can describe the variability behavior.
The value of $\sigma^{2}_{\rm n}$ (0.01 for the $V$-band LC; 0.04 for the X-ray LC) is larger than the squared of the mean size of the light curve uncertainties ($\overline{\sigma_y}^2$) where $\overline{\sigma_y}^2$=0.0001 for $V$-band and 0.0036 for X-ray data.
We have $\sigma^{2}_{\rm DRW}>$$\sigma^{2}_{\rm n}+$$\overline{\sigma_y}^2$ for both the $V$-band and X-ray data, 
which ensures that the fitted DRW amplitude is reasonable \citep{2021Sci...373..789B}.
It is possible that the quoted measurement errors are underestimated, and the excess white noise can account for excess measurement noise.

The $\gamma$-ray variability of the three sources has been analyzed in our previous work \citep{2022ApJ...930..157Z} with the same method.
We give the multi-band timescales with the errors in 95$\%$ confidence intervals of the three sources in Table~\ref{tab:X-optical-gamma tau}.
For 3C 273,  the $B$-band, X-ray, and $\gamma$-ray timescales are consistent within the errors.
The $V$-band and radio PSDs are single power-law, having no corresponding characteristic timescales.
For PKS 1510-089, the $V$-band, X-ray and $\gamma$-ray timescales are consistent within the errors but the $B$-band one has a smaller value.
For BL Lac, the $V$-band, X-ray and $\gamma$-ray timescales are also consistent within the errors.

\begin{deluxetable*}{cccccccc}
	\tablecaption{Modeling results of optical data for 38 blazars.\label{tab:fit}}
	\tablewidth{50pt}
	\setlength{\tabcolsep}{2mm}{
	\tablehead{
		\colhead{Object} & \colhead{Data sources} &\colhead{Waveband} &\multicolumn{2}{c}{Parameter of DRW} & \colhead{Damping timescale} &\colhead{Cadence} &\colhead{Length} \\
		\cmidrule(r){4-5}
		\colhead{} & \colhead{} & \colhead{} & \colhead{ln $\sigma_{\rm DRW}$} & \colhead{ln $\tau_{\rm DRW}$} & \colhead{(days)} & \colhead{(days)} & \colhead{(days)} \\
		\colhead{(1)} & \colhead{(2)} & \colhead{(3)} & \colhead{(4)} & \colhead{(5)} & \colhead{(6)} & \colhead{(7)} & \colhead{(8)}
	}
	\startdata
	1ES 1959+650 & SO & V & $-1.36^{+0.25}_{-0.17}$ & $5.13^{+0.53}_{-0.39}$ & $169^{+90}_{-66}$ & 23.5 & 3548.2 \\
	1ES 2344+514 & SO & V & $-3.14^{+0.20}_{-0.16}$ & $4.92^{+0.58}_{-0.46}$ & $137^{+79}_{-63}$ & 17.35 & 3539.1 \\
	3C 66A & SO & V & $-1.15^{+0.28}_{-0.17}$ & $5.35^{+0.57}_{-0.36}$ & $210^{+120}_{-75}$ & 9.04 & 3217.9 \\
	3C 454.3 & SO & V & $-0.81^{+0.10}_{-0.09}$ & $3.82^{+0.21}_{-0.18}$ & 
	$46^{+10}_{-8}$ & 5.97 & 3563.3 \\
	PKS 0235+164 & SO & V & $-0.41^{+0.13}_{-0.11}$ & $3.93^{+0.28}_{-0.24}$ & $51^{+14}_{-12}$ & 14.0 & 3417.9 \\
	4C +38.41 & SO & V & $-0.98^{+0.09}_{-0.08}$ & $3.51^{+0.19}_{-0.17}$ & $33^{+6}_{-6}$ & 9.7 & 3561.2 \\
	CTA 102 & SO & V & $-0.10^{+0.15}_{-0.12}$ & $4.26^{+0.32}_{-0.26}$ & $71^{+23}_{-18}$ & 10.13 & 3182.2 \\
	Mkn 421 & SO & V & $-1.19^{+0.19}_{-0.14}$ & $4.97^{+0.39}_{-0.29}$ & $144^{+56}_{-42}$ & 5.95 & 3562.7 \\
	Mkn 501 & SO & V & $-3.13^{+0.10}_{-0.09}$ & $3.92^{+0.24}_{-0.21}$ &
	$50^{+12}_{-11}$ & 7.31 & 3561.0 \\
	OJ 287 & SO & B & $-1.01^{+0.11}_{-0.09}$ & $3.57^{+0.23}_{-0.19}$ & $36^{+8}_{-7}$ & 5.32 & 3079.7 \\
	4C +21.35 & SO & V & $-1.46^{+0.18}_{-0.13}$ & $4.79^{+0.37}_{-0.28}$ & $120^{+45}_{-34}$ & 7.70 & 3357.9 \\
	PKS 2155-304 & SO & V & $-1.28^{+0.15}_{-0.12}$ & $4.42^{+0.32}_{-0.26}$ & $83^{+27}_{-22}$ & 11.03 & 3561.2 \\
	S5 0716+714 & SO & V & $-0.96^{+0.11}_{-0.09}$ & $2.99^{+0.24}_{-0.21}$ & $20^{+5}_{-4}$ & 14.98 & 3414.9 \\
	W Com & SO & V & $-1.13^{+0.18}_{-0.13}$ & $4.74^{+0.37}_{-0.29}$ & $114^{+42}_{-33}$ & 9.94 & 3538.7 \\
	4C +01.02 & S & B & $-1.53^{+0.18}_{-0.13}$ & $3.27^{+0.40}_{-0.32}$ & $26^{+11}_{-8}$ & 10.35 & 1408.2 \\
	PKS 0208-512 & S & V & $-0.56^{+0.15}_{-0.12}$ & $4.35^{+0.30}_{-0.24}$ & $77^{+23}_{-19}$ & 5.22 & 3301.9 \\
	PKS 0235-618 & S & R & $-0.97^{+0.12}_{-0.10}$ & $2.73^{+0.29}_{-0.25}$ & $15^{+4}_{-4}$ & 6.76 & 966.6 \\
	PKS 0402-362 & S & V & $-1.18^{+0.20}_{-0.15}$ & $3.91^{+0.42}_{-0.32}$ & $50^{+21}_{-16}$ & 8.39 & 1459.0 \\
	PKS 0426-380 & S & R & $-0.34^{+0.31}_{-0.19}$ & $4.85^{+0.64}_{-0.39}$ & $128^{+82}_{-50}$ & 2.63 & 1509.0 \\
	PKS 0458-02 & S & R & $-1.35^{+0.17}_{-0.13}$ & $3.20^{+0.38}_{-0.31}$ & $25^{+9}_{-8}$ & 10.53 & 1148.1 \\
	PKS 0502+049 & S & B & $-0.56^{+0.16}_{-0.13}$ & $2.80^{+0.36}_{-0.28}$ & $16^{+6}_{-5}$ & 5.87 & 769 \\
	PKS 0528+134 & S & V & $-1.38^{+0.20}_{-0.15}$ & $3.38^{+0.48}_{-0.38}$ & $29^{+14}_{-11}$ & 6.60 & 956.6 \\
	PMN J0531-4827 & S & V & $0.02^{+0.23}_{-0.17}$ & $4.03^{+0.49}_{-0.37}$ & $56^{+28}_{-21}$ & 9.27 & 1808.1 \\
	PMN J0850-1213 & S & R & $-0.79^{+0.14}_{-0.11}$ & $3.39^{+0.31}_{-0.27}$ & $30^{+9}_{-8}$ & 11.86 & 1696.6 \\
    PKS 1144-379 & S & B & $-0.62^{+0.30}_{-0.19}$ & $4.72^{+0.62}_{-0.41}$ & $112^{+70}_{-46}$ & 8.99 & 1590.7 \\
    PKS 1244-255 & S & R & $-0.99^{+0.12}_{-0.10}$ & $2.85^{+0.28}_{-0.25}$ & $17^{+5}_{-4}$ & 7.90 & 1515.9 \\
    PKS B1406-076 & S & R & $-1.19^{+0.10}_{-0.09}$ & $3.24^{+0.22}_{-0.20}$ & $26^{+6}_{-5}$ & 130.9 & 3365.9 \\
    PKS 1730-130 & S & V & $-1.59^{+0.14}_{-0.11}$ & $2.91^{+0.30}_{-0.25}$ & $18^{+6}_{-5}$ & 4.48 & 1008.2 \\
    PKS 1954-388 & S & R & $-1.19^{+0.16}_{-0.13}$ & $3.57^{+0.39}_{-0.34}$ & $36^{+14}_{-12}$ & 20.12 & 1851.0 \\
    PKS 2142-75 & S & V & $-2.03^{+0.11}_{-0.10}$ & $3.04^{+0.26}_{-0.22}$ & $21^{+5}_{-5}$ & 11.24 & 1843.1 \\
    PKS 2233-148 & S & R & $-0.30^{+0.17}_{-0.13}$ & $3.37^{+0.36}_{-0.28}$ & $29^{+10}_{-8}$ & 6.34 & 1110.2 \\
    PKS 2326-502 & S & B & $-0.38^{+0.22}_{-0.16}$ & $3.11^{+0.46}_{-0.34}$ & $22^{+10}_{-8}$ & 4.29 & 720.1 \\
    PMN J2345-1555 & S & R & $-0.48^{+0.18}_{-0.14}$ & $3.44^{+0.39}_{-0.30}$ & $31^{+12}_{-9}$ & 6.59 & 1424.1 \\
    Ton 599 & SO & V & $-0.39^{+0.20}_{-0.15}$ & $3.78^{+0.42}_{-0.32}$ & $44^{+18}_{-14}$ & 6.85 & 1143.9 \\
    PKS 2052-47 & S & V & $-0.91^{+0.22}_{-0.16}$ & $4.41^{+0.54}_{-0.40}$ & $82^{+44}_{-33}$ & 10.25 & 2121.2 \\
    3C 273 & SO & B & $-2.50^{+0.15}_{-0.12}$ & $4.08^{+0.31}_{-0.26}$ & $59^{+18}_{-15} $ & 8.9 & 3179.12 \\
    & SO & V & $-1.93^{+0.64}_{-0.46}$ & $8.06^{+1.29}_{-0.93}$ & $\cdots$ & 4.9 & 3423.6 \\
    PKS 1510-089 & SO & V & $-1.16^{+0.10}_{-0.09}$ & $3.67^{+0.22}_{-0.19}$ & $39^{+9}_{-7}$ & 8.91 & 3476.7 \\
	 & SO & B & $-1.26^{+0.06}_{-0.05}$ & $2.40^{+0.13}_{-0.12}$ & $11^{+1}_{-1}$ & 4.92 & 3315.9 \\
	BL Lac & SO & V & $-0.91^{+0.10}_{-0.08}$ & $3.86^{+0.25}_{-0.22}$ & $47^{+12}_{-10}$ & 4.78 & 3569.2 
	\enddata
	\tablecomments{ 
	(1) source name, (2) data source, S is for SMARTS and SO is for Steward Observatory blazar data archive, (3) wavebands of optical data, including B, V, R-band here, (4)(5) posterior parameters of modeling LC with DRW model, (6) damping timescale in the observed frame. The uncertainties of model parameters and damping timescales represent $1\sigma$ confidence intervals, (7) the mean cadence of the LC, and (8) the length of LC.
		}}
\end{deluxetable*}

\begin{deluxetable*}{ccccccc}
	\tablecaption{Modeling results of X-ray data for the 3 blazars.\label{tab:x-ray fit}}
	\tablewidth{50pt}
	\setlength{\tabcolsep}{4mm}{
	\tablehead{
		\colhead{Object} &\multicolumn{3}{c}{Parameter of DRW} & \colhead{Damping timescale} &\colhead{Cadence} &\colhead{Length} \\
		\cmidrule(r){2-4}
		\colhead{} & \colhead{ln $\sigma_{\rm DRW}$} & \colhead{ln $\tau_{\rm DRW}$} & \colhead{ln $\sigma_{\rm n}$} & \colhead{(days)} & \colhead{(days)} & \colhead{(days)} \\
		\colhead{(1)} & \colhead{(2)} & \colhead{(3)} & \colhead{(4)} & \colhead{(5)} & \colhead{(6)} & \colhead{(7)}
	}
	\startdata
	3C 273 & $0.91^{+0.06}_{-0.05}$ & $3.34^{+0.12}_{-0.11}$ & $\cdots$ & $28^{+3}_{-3}$ & 2.97 & 5811.5 \\
	PKS 1510-089 & $-1.87^{+0.05}_{-0.05}$ & $3.25\pm 0.12$ & $\cdots$ & $26^{+3}_{-3}$ & 4.16 & 5495.3 \\
	BL Lac & $-1.18^{+0.14}_{-0.11}$ & $4.15^{+0.33}_{-0.26}$ & $-1.64^{+0.03}_{-0.04}$ & $63^{+21}_{-16}$ & 2.19 & 2493.1 
	\enddata
	\tablecomments{ 
	(1) source name, (2)(3)(4) posterior parameters of modeling LC with DRW model, and (5) damping timescale in the observed frame. The uncertainties of model parameters and damping timescales represent $1\sigma$ confidence intervals, (6) the mean cadence of the LC, and (7) the length of LC.
		}}
\end{deluxetable*}

\begin{deluxetable*}{ccccc}
	\tablecaption{Damping timescale of 3C 273, PKS 1510-089 and BL Lac.\label{tab:X-optical-gamma tau}}
	\setlength{\tabcolsep}{4mm}{
	\tablehead{
		\colhead{Object} & \colhead{$B$-band timescale} & \colhead{$V$-band timescale} & \colhead{X-ray timescale} & \colhead{$\gamma$-ray timescale} \\
		\colhead{} & \colhead{(days)} & \colhead{(days)} & \colhead{(days)} & \colhead{(days)} \\
		\colhead{(1)} & \colhead{(2)} & \colhead{(3)} & \colhead{(4)} & \colhead{(5)}
	}
	\startdata
	3C 273 & $59^{+41}_{-28}$ & unreliable & $28^{+7}_{-6}$ & $31^{+12}_{-10}$  \\
	PKS 1510-089 & $11^{+3}_{-3}$ & $39^{+18}_{-14}$ & $26^{+7}_{-6}$ & $40^{+14}_{-12}$ \\
	BL Lac & no data & $47^{+26}_{-19}$ & $63^{+49}_{-30}$ & $69^{+36}_{-25}$ \\
    \enddata
	\tablecomments{ 
	(1) source name, (2)(3)(4)(5) multi-band damping timescales in the observed frame. The uncertainties of the damping timescales represent 95$\%$ confidence intervals of the distribution of the parameter.
		}}
\end{deluxetable*}

\subsection{Optical Results of 38 Blazars} \label{subsec:large samples}

The DRW model can successfully fit the long-term optical LCs of the 38 blazars.
Based on the criteria of selecting reliable measurements of the damping timescale, 
we get reliable optical timescales for the 38 blazars.
The basic information of the 38 blazars and the modeling results are given in Table~\ref{tab:information} and Table~\ref{tab:fit}, respectively.
Except for 3C 273 and PKS 1510-089 which are analyzed in Section \ref{subsec:individual}, 
the timescales for different optical bands are consistent for the other 36 sources. 
This indicates that the optical emission of the 36 blazars has the same origin, i.e., the jet emission.
In Table~\ref{tab:fit}, we only list one optical band result for these sources.  
The timescale is between 10 days and 200 days.

Some notes should be given on PKS 2052-47 and Ton 599.
The fitting to the LC of PKS 2052-47 needs an additional white noise, 
and the relation $\sigma^{2}_{\rm DRW} (0.16)>\sigma^{2}_{\rm n} (0.026)+\overline{\sigma_y}^2 (0.0016)$ still holds.
Ton 599 has big gaps and few data in the first half of its $V$-band LC, and we select the second half of the LC to analyze.

\subsection{Origin of the optical emission from 3C 273 and PKS 1510-089} \label{subsec:originOptical}

The optical emission of 3C 273 and PKS 1510-089 is complicated.
Blue bump can be seen in their multi-band spectral energy distributions \citep[SEDs; e.g.,][]{2010ApJ...716...30A,2012ApJ...760...69N,2017A&A...601A..30C}.
SED modeling results showed that the accretion disk has a significant contribution to the optical emissions of 3C 273 and PKS 1510-089 \citep[e.g.,][]{2012ApJ...760...69N,2012PASJ...64...80Y,2017A&A...601A..30C}.
In addition, \cite{2019ApJ...876...49Z} found that a long-term variation trend in the optical continuum LC of 3C 273 does not appear in the emission-line variation.
This suggests that the long-term variation trend is not contributed by the accretion disk, and it could originate from the jet.
\cite{2020ApJ...897...18L} quantitatively decoupled the optical emissions from the jet and accretion
disk in 3C 273 and found that the jet emission accounts for 10\%-40\% of the total optical emission.
\citet{2022MNRAS.510.1809P} studied the correlation between $V$-band flux and polarization degree (PD) variations using SO observation during 2008-2018.
They found a significant positive correlation only in two of the ten observing cycles.
Note that the PD is quite small, and it changes from 0.04\% to 1.58\% during 2008-2018.
The $V$-band single power-law PSD we obtained here is different from the typical PSD of the accretion disk \citep{2021ApJ...907...96S,2021Sci...373..789B} 
and jet variability \citep{2022ApJ...930..157Z}.
The complicated mixture of the jet and accretion disk emissions at the $V$-band may result in the single power-law PSD.
The mixed emission also results in the weak correlation between $V$-band and {\it Fermi} $\gamma$-ray variabilities reported by \citet{2021ApJ...923....7B}.
We find no significant correlation between $B$-band variability and $\gamma$-ray variability for 3C 273 and PKS 1510-089.
Looking at the location of the blue bump in SED \citep{2021MNRAS.504.1103R}, we suggest that the $B$-band emission of 3C 273 is dominated by the accretion disk photons.

For PKS 1510-089, the $V$ and $B$-band timescales are clearly different, indicating different origins for the two bands' emissions.
The $V$-band polarization of PSK 1510-089 is averagely greater than that of 3C 273, varying from 0.2\% to 25.82\% \citep{2022MNRAS.510.1809P}.
Among the ten observing cycles during 2008-2018, a significant positive correlation between $V$-band flux and PD variations is found in 5 cycles.
Moreover, \citet{2017A&A...601A..30C} found a good correlation between the long-term SO $V$-band and $\gamma$-ray LCs.
These results suggest that the $V$-band emission is dominated by jet contribution.
Also looking at the location of the blue bump in SED \citep{2012ApJ...760...69N}, 
the $B$-band emission with a smaller timescale of 11 days is suggested as the accretion disk contribution.

\subsection{Comparing Optical and $\gamma$-ray results} \label{subsec:ComOptical}

\begin{figure}
    \centering
    {\includegraphics[width=10cm]{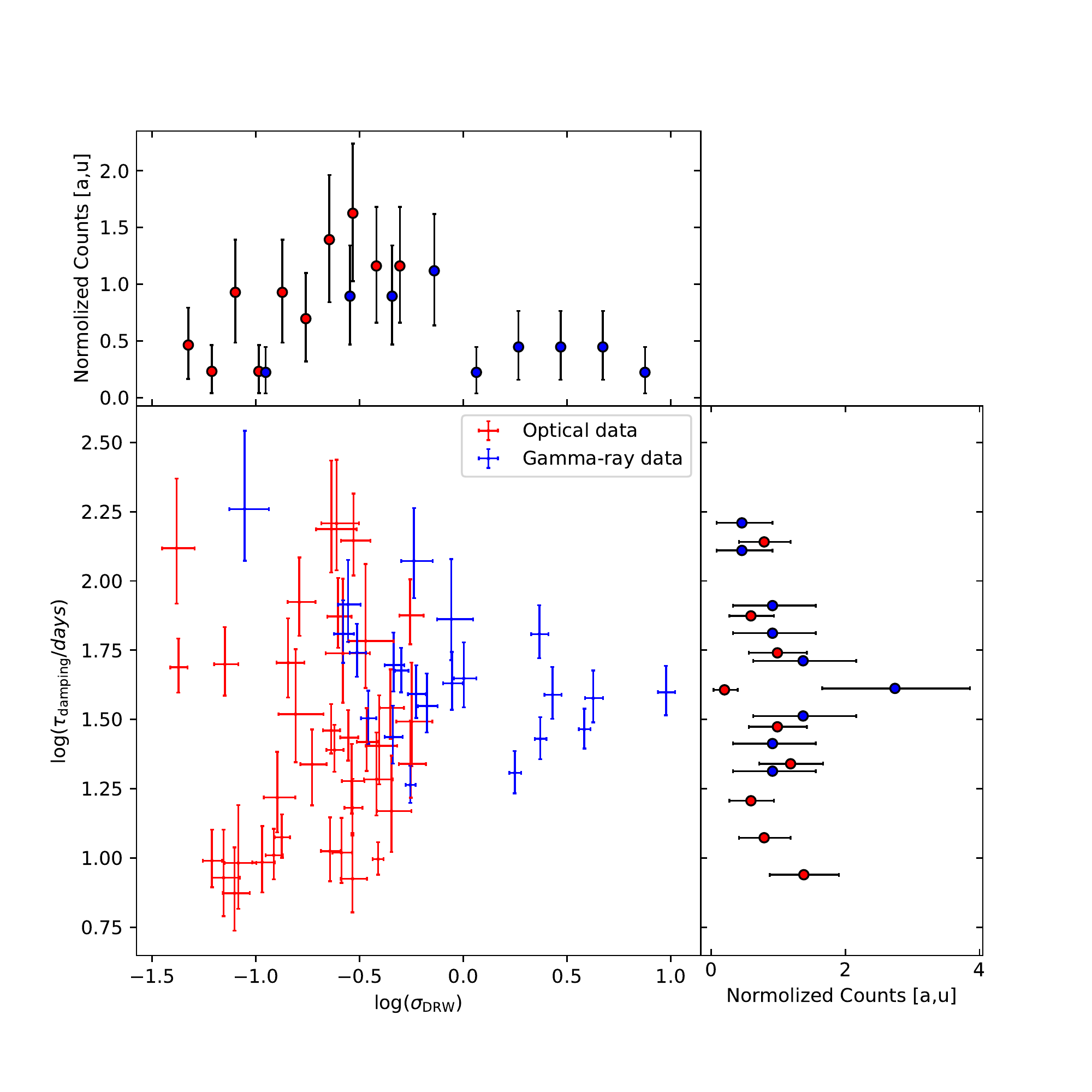}}
    \caption{
    Plot of the redshift-corrected timescale $\tau_{\rm DRW}$ versus the amplitude \textbf{$\sigma_{\rm DRW}$}.
    The red and blue points represent the optical and $\gamma$-ray results, respectively.
    The side panels show the normalized histograms of the distributions of redshift-corrected $\tau_{\rm DRW}$ (right) and $\sigma_{\rm DRW}$ (top) for blazars.
\label{fig:tau-sigma}}
\end{figure}

Long-term {\it Fermi} $\gamma$-ray LCs of 22 blazars have been analyzed by \cite{2022ApJ...930..157Z} with the same GP method.
The optical timescale in this work is generally consistent with the $\gamma$-ray timescale (Figure~\ref{fig:tau-sigma}).
We examine the consistency of the timescales in the two energy-bands by using a statistical significance test (T-test).
We get t-statistic=1.1 and $p$-value=0.28 ($\textgreater$0.05), 
which means that in statistic there is little difference between the two groups of timescales.
The optical amplitude term $\sigma_{\rm DRW}$ is less than one, and the $\gamma$-ray $\sigma_{\rm DRW}$ can be greater than 10.
This means that $\gamma$-ray variability can be more energetic than optical variability.

We separated the sources into two groups with $\rm M_{\rm BH}<10^{9}M_{\rm \odot}$ and $\rm M_{\rm BH}>10^{9}M_{\rm \odot}$.
The mean timescales (redshift-corrected) in different ranges of black hole mass are listed in Table~\ref{tab:mean-timescale}.
It is found that the mean timescale of the sources in the mass range of $10^{9}$-$10^{10}\rm M_{\rm \odot}$ is smaller
in both $\gamma$-ray and optical energies.
However, we have a  few sources with the mass of $10^{9}$-$10^{10}\rm M_{\rm \odot}$, 
therefore this result may be tentative.

\begin{figure}
    \centering
    {\includegraphics[width=10cm]{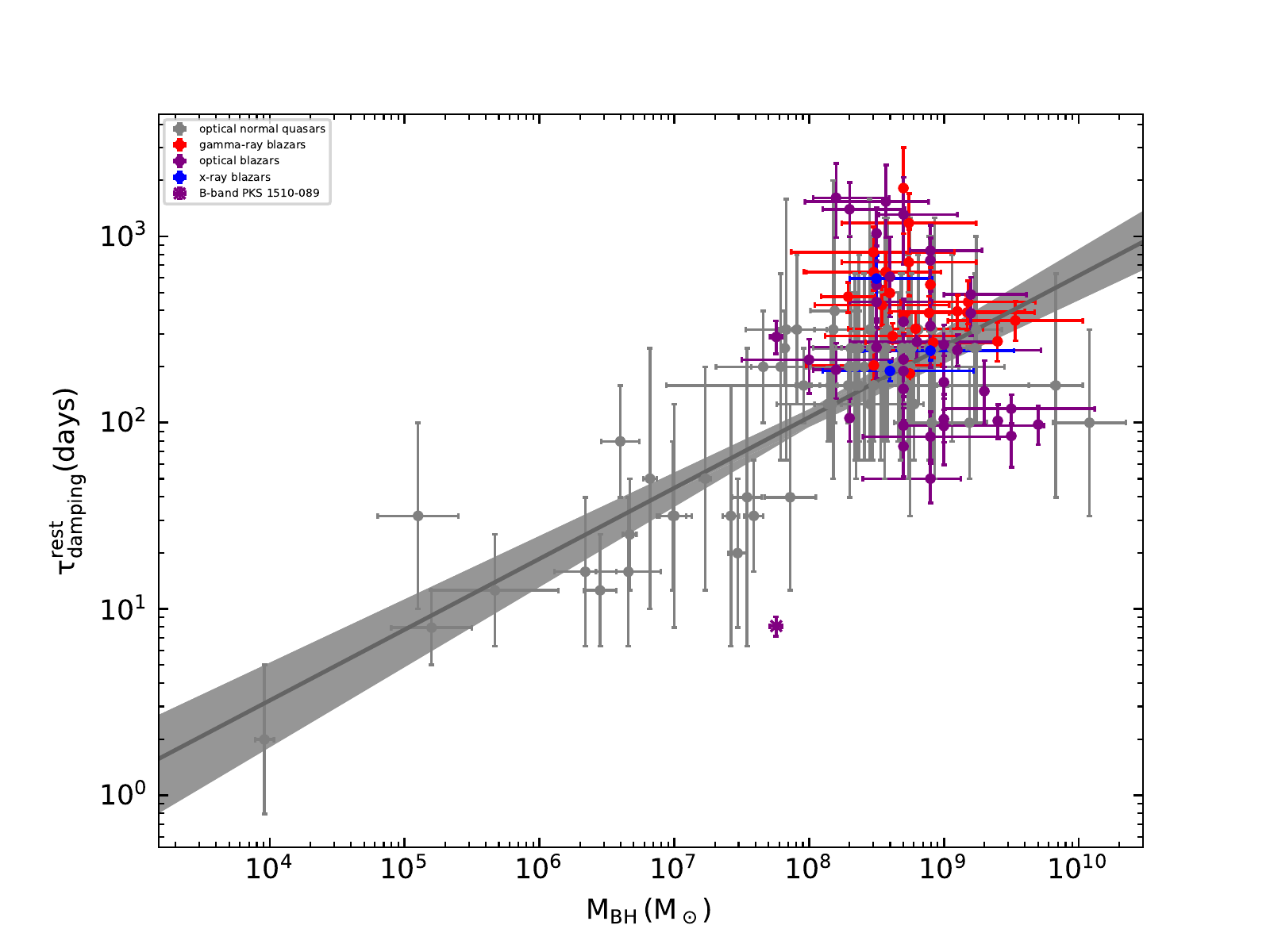}}
    \caption{Plot of the rest-frame timescale versus black hole mass.
    The gray data, lines, and area represent the optical accretion disk results for normal quasars taken from \cite{2021Sci...373..789B}. Red data are $\gamma$-ray results for blazars taken from \cite{2022ApJ...930..157Z}, 
    and the purple and blue data respectively represent the optical and X-ray results for blazars obtained in this work. 
\label{fig:tau-mass}}
\end{figure}

In Figure~\ref{fig:tau-mass}, we plot the relationship between the damping timescale in the rest frame ($\tau^{\rm rest}_{\rm damping}$) and the black hole mass of blazars 
along with the results of normal quasars from \cite{2021Sci...373..789B}.
The timescales should be modified into the rest frame with the following formula:
\begin{equation}\label{eq4}
\tau^{\rm rest}_{\rm damping}=\frac{\tau_{\rm DRW}\ \delta_{\rm D}}{1+z}\;.
\end{equation}
An average Doppler factor of $\delta_{\rm D}$=10 is used here and the redshift $z$ for each source is given in table~\ref{tab:information}.
We show the optical, X-ray, and $\gamma$-ray results in the plot.
It is found that the nonthermal optical $\tau^{\rm rest}_{\rm damping}$ of blazars and the thermal optical timescale of normal quasars occupy the same space in the plot of $\tau^{\rm rest}_{\rm damping}-\rm M_{\rm BH}$.

The X-ray results for the three individual blazars are also in the same area as the optical results.
The $B$-band timescale of 3C 273 is a typical value of accretion disk timescale.
The $B$-band timescale of  PKS 1510-089 is an outlier value among the accretion disk timescales.
This value significantly deviates from the relation between damping timescale and black hole mass reported by \cite{2021Sci...373..789B}.

\begin{deluxetable*}{ccccccc}
	\tablecaption{Mean timescales (redshift-corrected) of blazars in $\gamma$-ray and optical energies.\label{tab:mean-timescale}}
	\tablewidth{50pt}
	\setlength{\tabcolsep}{4mm}{
	\tablehead{
		\colhead{Waveband} & \colhead{log$\rm M_{\rm BH}/\rm M_{\rm \odot}$} &\colhead{Mean timescale} \\
		\colhead{(1)} & \colhead{(2)} & \colhead{(3)}
	}
	\startdata
	$\gamma$-ray & $8-9$ & $58^{+21}_{-16}$ \\
	 & $9-10$ & $32^{+10}_{-8}$ \\
	 & $8-10$ & $53^{+18}_{-14}$ \\
	optical & $8-9$ & $51^{+23}_{-11}$ \\
	& $9-10$ & $19^{+6}_{-5}$ \\
	 & $8-10$ & $42^{+18}_{-13}$
	\enddata
	\tablecomments{ 
	(1) waveband, (2) the range of black hole mass in solar mass,  (3) the mean damping timescale (redshift-corrected) with unit day. The uncertainties of timescales represent $1\sigma$ confidence intervals. 
		}}
\end{deluxetable*}

\section{Discussion} \label{sec:discussion}
It is difficult to directly resolve the inner jet structure of the blazar\footnote{The inner parsec jet of the blazar J1924–2914 has been resolved by the Event Horizon Telescope \citep{2022ApJ...934..145I}.}.
Especially, the location of the high-energy emission region is still a hot open question \citep[e.g.,][]{2016ARA&A..54..725M,2019Galax...7...20B}.
Multi-band variability analysis provides an indirect approach to resolve the emission regions.
The cross-correlation method is frequently used in multi-band variability analysis \citep[e.g.,][]{2018MNRAS.480.5517L,2021ApJ...923....7B}.

GP method has been wildly used to characterize the AGN accretion disk variability \citep{2009ApJ...698..895K,2018ApJ...853..116Z,2019ApJ...877...23L,2021Sci...373..789B}.
In blazar science, it becomes popular in recent several years \citep[e.g.,][]{2018ApJ...863..175G,2019ApJ...885...12R,2020ApJ...895..122C,2020ApJS..250....1T,2021ApJ...907..105Y,2022ApJ...930..157Z}.
In this work, we use the GP method to study the multi-band variability of the blazar.
This provides results independent of the cross-correlation method.

The $\gamma$-ray variability of the blazar has been studied by \citet{2022ApJ...930..157Z} with the GP method.
Here we focus on the X-ray and optical variability of the blazar.
Multi-band emission from the blazar is dominated by the nonthermal jet contribution.
Two special blazars are 3C 273 and PKS 1510-089. 
An optical-ultraviolet bump appears in their SED, which is associated with their thermal accretion disk emission \citep[e.g.,][]{2012ApJ...760...69N,2012PASJ...64...80Y,2017A&A...601A..30C}.

We fit the long-term optical LCs from the database of SO and SMARTS with the DRW model.
Finally, 38 blazars with a reliable characteristic timescale are selected.
Except for 3C 273 and PKS 1510-089, the timescales in different optical colors agree with each other for the remaining 36 blazars.
This indicates that the emissions in different optical colors of the 36 blazars have the same origin, i.e., the jet emission.

\cite{2012ApJ...760...51R} modeled the optical LCs covering from 2002 December through 2008 March of 51 blazars using the DRW model.
They found that the observed damping timescale peaks at $\sim$80 days, 
and the intrinsic timescale $\tau_{\rm damping}^{\rm rest}$ peaks at $\sim$800 days\footnote{They also used $\delta_{\rm D}=10$ for the Doppler effect correction.}.
The distribution of the optical timescale obtained in this work is flat (Figure~\ref{fig:tau-sigma}), 
and the average optical $\tau_{\rm damping}^{\rm rest}$ is $\sim$400 days, which is smaller than the result of \cite{2012ApJ...760...51R}.
All blazars in our sample are Fermi-detected $\gamma$-ray sources. 
While the sample studied by \cite{2012ApJ...760...51R} would be dominated by the blazars of non-Fermi detection.
Therefore, the results indicate that the optical timescale of the blazar of non-Fermi detection may be longer than that of the blazar of Fermi detection.
\citet{2015MNRAS.450.3568X} found that the two population blazars indeed have different physical properties, for example, the blazar of non-Fermi detection has a smaller Doppler factor \citep{2017ApJ...851...33P}.

In the reverberation mapping studies of 3C 273 and PSK 1510-089, a nonechoed long-term trend is found in the optical continuum LC \citep{2019ApJ...876...49Z,2020ApJ...897...18L,2020AA...642A..59R}.
This reveals the mixed origin of their optical emission.
New clues on the origin of the optical emission can be found in our results.
The $V$ and $B$-band timescales of PSK 1510-089 are different.
Its long-term $V$-band variability is correlated with the $\gamma$-ray variability \citep{2017A&A...601A..30C}, suggesting that the $V$-band emission is dominated by jet contribution.
The long-term polarization variation \citep{2022MNRAS.510.1809P} also supports 
 that the nonthermal component is dominated at $V$-band.
 The $V$-band emission of 3C 273 seems to be more complicated.
 The jet contribution to $V$-band emission may be strongly time-dependent and may vary in a large range.
 This complicated mixture of jet and accretion disk emission results in a single power-law PSD.
 For the two sources, no significant correlation is found between $B$-band and $\gamma$-ray variabilities in our analysis.
  The $B$-band emission is naturally considered as the accretion disk contribution.
For 3C 273, the $B$-band timescale of $\approx60$ days is a typical value for the accretion disk emission of normal quasars.
While the $B$-band timescale of $\approx11$ days of PKS 1510-089 is significantly smaller, 
and it deviates from the $\tau^{\rm rest}_{\rm damping}-\rm M_{\rm BH}$ relation of \cite{2021Sci...373..789B} (Figure~\ref{fig:tau-mass}).
This short timescale may imply special properties of its accretion disk.

The nonthermal optical, X-ray and $\gamma$-ray variabilities all have the typical DRW PSD.
Namely, the PSD of synchrotron emission is the same as that of inverse-Compton (IC) emission, 
consistent with the simulations with a time-dependent one-zone leptonic blazar emission model \citep{2022ApJ...925..177T}.
In other words, the long-term jet variability is irrelevant to the underlying emission mechanism.

\citet{2021Sci...373..789B} suggested that the DRW damping timescale measured from the accretion disk variability of normal quasars could be associated with the thermal instability timescale expected in the AGN standard accretion disk theory. 
\citet{2022ApJ...930..157Z} measured the $\gamma$-ray DRW damping timescale of AGNs from the Fermi-LAT data, and found that the $\gamma$-ray timescales of 23 AGNs
occupy almost the same space with the optical variability timescales of normal quasars in the plot of $\tau^{\rm rest}_{\rm damping}-\rm M_{\rm BH}$.
In this work, we add the nonthermal optical timescale of blazars in this plot.
The nonthermal optical timescale of blazars also locates at the same region with the thermal optical timescale of normal quasars in the plot (Figure~\ref{fig:tau-mass}). 
This implies that the jet variability is relevant to the accretion disk.
The thermal instability in accretion disk may not only cause the accretion disk variability but also the jet multi-band variability.

Statistically, the nonthermal optical $\tau_{\rm damping}^{\rm rest}$ of 38 blazars are consistent with the $\gamma$-ray $\tau_{\rm damping}^{\rm rest}$ of 22 blazars.
Individually (3C 273, PKS 1510-089, and BL Lac), the damping timescales of the jet variability in optical, X-ray, and $\gamma$-ray energies are consistent within the measured errors.
Our results indicate that multi-band jet emissions are produced in the same region.
However, we still cannot know the distance from the emission region to the central black hole.
The radio observation is helpful to constrain this distance \citep{2014MNRAS.445..428M}.
We modeled the OVRO radio LCs covering over $\sim$ten years, and we obtain a single power-law PSD. In this work, we only show the radio result for 3C 273 as an example.
We also modeled the 30-yr radio LCs of 3C 279 and 3C 454.3 obtained from Aalto University Mets$\ddot{a}$hovi Radio Observatory, and we still get an unconstrained timescale.
The results indicate the radio timescale is very large and may be larger than 10 years.
Through the very long baseline interferometry (VLBI) observation, one can determine the distance from the radio core to the central black hole.
Comparing the optical/X-ray/$\gamma$-ray timescale and the radio timescale, 
we can infer that the optical/X-ray/$\gamma$-ray emission region is far upstream from the radio core.

\section{Summary} \label{sec:summary}

We analyze the blazar's radio, optical, and X-ray variabilities using the GP tool {\it celerite}.
The DRW model can successfully fit the jet multi-band variabilities.
The multi-band characteristic timescale is used to probe the structure of the emission region in the blazar jet.
Our main results are as follows. 

$(\romannumeral1)$ 
The synchrotron and IC emissions have the same PSD, i.e., the typical DRW PSD.
This indicates that the jet's long-term variability is irrelevant to the underlying emission processes.
In the plot of $\tau^{\rm rest}_{\rm damping}-\rm M_{\rm BH}$, the jet timescales locate at almost the same space as the accretion disk timescales of normal quasars, 
 implying that the jet and accretion disk variability is driven by the same physical process \citep{2022ApJ...930..157Z}.

$(\romannumeral2)$ 
The nonthermal optical, X-ray, and $\gamma$-ray variability has a consistent characteristic timescale.
The radio characteristic timescale is very long which cannot be constrained by decades-long LC.
The results indicate that the nonthermal optical-X-ray-$\gamma$-ray emission is produced in the same region, which is upstream and far from the radio core. This supports the basic hypothesis of the standard Synchrotron-Self-Compton jet model.

The GP method provides a flexible approach 
to understand the variability pattern of AGN in the framework of stochastic process.
Adopting the standard GP tool \citep{2017AJ....154..220F},
we build the link between accretion disk (thermal emission) and the jet (nonthermal emission), i.e., Figure~\ref{fig:tau-mass}.
This is a new methodology for comparing thermal and nonthermal emissions, additional to the comparison between the thermal and 
nonthermal luminosities \citep[e.g,][]{2011MNRAS.414.2674G,2012MNRAS.421.1764S,2014Natur.515..376G}.

\acknowledgments
We thank the referees' valuable report. This work is partially supported by the National Key R \& D Program of China under grant No. 2018YFA0404204.
H. Y. Zhang acknowledges the financial support from the Scientific Research Fund project of Yunnan Education Department (2022Y053) and the Graduate Research innovation project of Yunnan University (2021Y034).
The work of D. H. Yan is also supported by the CAS Youth
Innovation Promotion Association and Basic research Program of Yunnan Province (202001AW070013).

Data from the Steward Observatory spectropolarimetric monitoring project were used. This program is supported by Fermi Guest Investigator grants NNX08AW56G, NNX09AU10G, NNX12AO93G, and NNX15AU81G. 
This research has made use of up-to-date SMARTS optical/nearinfrared light curves.
This research has made use of data from the OVRO 40-m monitoring program, which is supported by private funding from the California Insitute of Technology and the Max Planck Institute for Radio Astronomy, and by NASA grants NNX08AW31G, NNX11A043G, and NNX14AQ89G and NSF grants AST-0808050 and AST- 1109911.
This work also has made use of $\left\{\rm lightcurves\right\}\left\{\rm spectral\; files\right\}$ provided by the University of California, San Diego Center for Astrophysics and Space Sciences, X-ray Group (R.E. Rothschild, A.G. Markowitz, E.S. Rivers, and B.A. McKim).

{\it Facility:} SMARTS.

{\it Software:} corner.py \citep{2016JOSS....1...24F}, celerite \citep{2017AJ....154..220F}, emcee \citep{2013PASP..125..306F}, NumPy \citep{2020NumPy-Array}, Matplotlib \citep{2007CSE.....9...90H}, Astropy \citep{2013A&A...558A..33A,2018AJ....156..123A}, SciPy \citep{2020SciPy-NMeth}.

\bibliography{ms.bib}{}
\bibliographystyle{aasjournal}

\end{CJK*}
\end{document}